\documentclass[sii]{ipart}

\usepackage{amsthm,amsmath}
\usepackage[numbers,square]{natbib}
\RequirePackage{hyperref}
\usepackage{amssymb}
\usepackage{mathrsfs}
\usepackage{algorithmic}
\usepackage{algorithm}
\usepackage{graphicx}
\usepackage{caption}
\usepackage{subcaption}
\usepackage{threeparttable}
\usepackage{tabularx}
\usepackage{mathtools}

\pubyear{0000}
\volume{0}
\issue{0}
\arxiv{1412.1559}

\startlocaldefs
\newcommand{\calA}{\mathcal{A}}
\newcommand{\defi}{\mathop{=}\limits^{\Delta}}      
\newcommand{\algorithmicbreak}{\textbf{break}}
\newcommand{\BREAK}{\STATE \algorithmicbreak}

\DeclareMathOperator*{\argmax}{arg\!\max}
\endlocaldefs

\begin{document}

\begin{frontmatter}

\title{Iterative Subsampling in Solution Path Clustering of Noisy Big Data}
\runtitle{ISSPC of Big Data}


\author{\fnms{Yuliya} \snm{Marchetti}\thanksref{t1}\ead[label=e1]{yuliya@stat.ucla.edu}}
\address{Department of Statistics \\ University of California, Los Angeles \\ 8125 Math Sciences Bldg. \\ 
Los Angeles, CA 90095 \\ \printead{e1}}
\and
\author{\fnms{Qing} \snm{Zhou}\thanksref{t2,t3}\ead[label=e2]{zhou@stat.ucla.edu}}
\address{Department of Statistics \\ University of California, Los Angeles \\ 8125 Math Sciences Bldg. \\ 
Los Angeles, CA 90095 \\ \printead{e2}}

\thankstext{t1}{Supported by a UCLA dissertation year fellowship.}
\thankstext{t2}{Supported in part by NSF grants DMS-1055286 and DMS-1308376.}
\thankstext{t3}{Corresponding author.}

\runauthor{Y. Marchetti and Q. Zhou}

\begin{abstract}
We develop an iterative subsampling approach to improve the computational efficiency of our previous work
on solution path clustering (SPC). The SPC method achieves 
clustering by concave regularization on the pairwise distances between cluster centers. 
This clustering method has the important 
capability to recognize noise and to provide a short path of clustering solutions; however,
it is not sufficiently fast for big datasets. Thus, we propose a method that iterates between clustering a small subsample of the full data and sequentially 
assigning the other data points to attain orders of magnitude of 
computational savings. The new method preserves the ability to isolate noise, includes a 
solution selection mechanism that ultimately provides one clustering solution with an estimated number of 
clusters, and is shown to be able to extract small tight clusters from noisy data. 
The method's relatively minor losses in accuracy are demonstrated through simulation studies, and its ability to handle large 
datasets is illustrated through applications to gene expression datasets.  
An R package, $\verb+SPClustering+$, for the SPC method with iterative subsampling is available at
\url{http://www.stat.ucla.edu/~zhou/Software.html}. 
\end{abstract}





\end{frontmatter}

\section{Introduction}

There have been a large number of sophisticated and state-of-the-art clustering methods developed in statistics and machine learning over the last few decades, in response 
to the rapid emergence of more complex and richer datasets. The exponential growth in stored data has ushered in the era of ``big data" and 
the interest to understand and exploit such information. The ever-increasing problem size and computational intensity of such efforts are further creating 
challenges for conventional data analysis methods. Large datasets are often noisy and consist of heterogeneous subsets. 
Consequently, clustering is an essential exploratory analysis to partition the data into smaller subsets and to filter out noisy
data points so that simpler and robust models may be constructed for each cluster. 

The challenges of clustering large datasets have been widely studied, and numerous methods have been proposed to efficiently handle such data. The majority of such methods, originating in machine learning and data mining applications, modify the popular k-means or EM algorithms in order to increase their computational speeds. Some examples of such modifications include the use of simple random subsampling or more sophisticated sampling schemes \cite{bradley1998, fayyad1998, guha2001, rocke2003, hathaway2006, ma2006, ma2014, boutsidis2015}, the computation of some summary representations of the data \cite{zhang1996, ng2002, posse2001}, and the parallelizing and distributing of the computation process \cite{ranger2007, ene2011}. A number of reviews of such techniques summarize and categorize the abundance of these methodologies \cite{aggarwal2013, jain2010,shirkhorshidi2014}.
 
More advanced clustering methodologies, on the other hand, make it possible, for example, to automatically estimate the number of clusters \cite{fraley2002, tseng2007}, to isolate outliers and noisy data points \cite{fraley2002, tseng2005, tseng2007}, i.e. data points showing no grouping pattern, to perform dimension and variable selection \cite{pan2007, zhou2009}, and to handle non-convex clusters \cite{hocking2011, chi2013, pan2013}. Although not as fast as the k-means or the EM algorithm, 
these methods are powerful, and it is very useful to develop computational strategies to allow the application of these clustering methodologies to moderately large datasets that otherwise would be prohibitively slow or simply impossible to manipulate. 

Perhaps, the simplest approach to clustering a large dataset is based on random subsampling, which would be computationally efficient if the subsample is very small, say with size on the order of $\sqrt{n}$, where $n$ is the sample size of a dataset. Subsampling for large datasets is not new and was first proposed in \cite{kaufman1990}. Banfield and Raftery \cite{banfield1993} cluster a real dataset using model-based clustering (mclust) and apply discriminant analysis to classify the remaining data points. Fayyad and Smyth \cite{fayyad1996} suggest iteratively subsampling datasets for clustering and classifying the remaining data points until all of them belong to the clusters identified in the subsamples with sufficiently high probability. Later, Fraley and Raftery \cite{fraley2002} elaborate on subsample clustering and discriminant analysis for large data and discuss a modification of the simple random subsampling with the goal of finding small, tight clusters. A number of other clustering methods were subsequently developed, following a similar idea \cite{maitra2001, wehrens2004, fraley2005, koestler2010, manolopoulou2010, naim2014}. All of these methods are geared mainly towards computational efficiency, and several were also developed to find small clusters in large datasets \cite{fayyad1996, maitra2001, fraley2005, manolopoulou2010, naim2014}. 

In this paper we propose a new algorithm, primarily motivated by the need to accelerate the solution path clustering (SPC) method developed in our recent work \cite{marchetti2014}. SPC produces a small set of clustering solutions, called a solution path, that includes not only cluster assignments but also an estimated decreasing number of clusters along the path. SPC is based on sparsity regularization and introduces a concave penalty to a quadratic loss function, which is minimized with penalty parameters chosen by a data-driven approach. The regularization on the pairwise distances between the cluster centers effectively makes it possible to achieve clustering and at the same time eliminates the need to specify the number of clusters as an input parameter. SPC is able to find compact clusters and detect noise as singleton or small clusters, due to the fact that initial clustering solutions are obtained by penalizing relatively small pairwise distances between the cluster centers. SPC, however, has time complexity $O(n^2)$ as it depends on the calculation of the pairwise distances between data points. 
To achieve considerable computational savings we introduce an iterative approach that enables applications of SPC to large datasets. For the new algorithm, we combine SPC, performed on a small subsample of the data, with subsequent assignment of the remaining data points based on likelihood ratio evaluation. We then iterate between the clustering and sequential assignment steps until no more valid clusters are found. The iterative subsampling SPC (ISSPC) provides orders of magnitude computational savings compared to the original SPC algorithm with relatively little loss in accuracy of the resulting clustering partition. It maintains the effectiveness of SPC to separate the noise from the clusters, eliminating the need for any prior filtering of the data. It can also be successful in locating small tight clusters in large datasets. Moreover, the new iterative subsampling approach utilizes SPC's solution path to obtain one final clustering solution with an estimated number of clusters, and in effect, performs fast and efficient solution selection.

We summarize the novel contributions of our work in comparison to existing subsampling-based clustering methods.  First, while all of the existing 
methods assume that the number of clusters is given or require that some initial estimate of the number of clusters is provided by the user 
\cite{fayyad1996, maitra2001, naim2014, banfield1993, fraley2002, wehrens2004, fraley2005}, ISSPC determines the number of clusters on its 
own through significance tests on a solution path.  This improves usability and at the same time reduces user bias. Second, both the clustering 
step and the assignment step in our approach are designed under the assumption that the full dataset may contain a certain proportion of noisy data points.  
In each iteration, only a portion of the data points are partitioned into clusters, and the remaining data points will be considered in the subsequent iterations 
until no more clusters are identified. On the contrary, all but one of the methods reviewed above partition the entire 
dataset into clusters generated from the subsample. The performance on noisy data has been demonstrated solely in \cite{wehrens2004} with data 
containing only $5\%$ of noise, while we show results for varying noise proportions up to $90\%$. Third, as most clustering algorithms including SPC 
have time complexity of $O(n^2)$, the subsample size we consider here is $O(\sqrt{n})$, which is much smaller than the subsample sizes used in the previous work \cite{banfield1993, fraley2002, wehrens2004, fraley2005, koestler2010, manolopoulou2010, naim2014}. This is important in the context of big data applications and inherently large datasets, for which only algorithms with $O(n)$ operations would 
be computationally feasible. We study systematically the performance loss and computational savings by varying the subsample size as 
$a\sqrt{n}$ for $a \in [1,10]$. Lastly, integrating SPC, a regularization clustering method, with subsampling raises a number of new challenges, 
such as the choice of tuning parameters in the concave penalty and model selection along the regularization path. This work provides practical solutions 
to these problems with satisfactory performance on both simulated and real data.

The remainder of this paper is organized as follows.
In Section~\ref{sec:spc} we provide a short review of SPC. The sequential assignment based on likelihood ratio calculation is described in Section~\ref{sec:sequential},
 and the full algorithm is presented in Section~\ref{sec:subsampleSPC}. Section~\ref{sec:tuning} discusses some practical considerations for choosing tuning parameters. 
 We demonstrate the performance of ISSPC on simulated data in Section~\ref{sec:simstudies} and on gene expression data in Section~\ref{sec:genedata}.
The article is concluded with a brief discussion.

\section{Methods}
\label{sec:methods}

\subsection{Solution path clustering}
\label{sec:spc}

Denote by $Y = (y_{im})_{n \times p}$ an observed data matrix, where $y_i = (y_{i1}, \ldots, y_{ip}) \in \mathbb{R}^p$ represents the $i$th object. Denote the cluster center for $y_i$ by $\theta_i \in \mathbb{R}^p$ for $i=1,\ldots,n$.
SPC minimizes a penalized $\ell_2$ loss function with a penalty $\rho(\cdot)$ imposed on the Euclidean distance between pairwise center parameters $\| \theta_i - \theta_{j} \|_2$:
\begin{align}
	\label{eq:kloss}
	& \ell(\theta_1, \ldots, \theta_n) = \sum_{i=1}^n\| y_i - \theta_i \|_2^2 + \lambda \sum_{ i < j } \rho \left( \| \theta_i - \theta_j \|_2 \right).
\end{align}
For $\rho(\cdot)$ we choose the minimax concave penalty (MCP) developed in \cite{zhang2010},
\begin{align}
	\label{eq:penalty}
	 \rho \left( t \right) & = \int_0^{t} \left( 1 - \frac{x}{\delta \lambda}  \right)_+ dx \\
	& = \left( t - \frac{t^2}{ 2 \lambda \delta }  \right) I( t < \lambda \delta ) + \left( \frac{ \lambda \delta }{ 2 } \right) I( t \geq \lambda \delta ) \notag
\end{align}
for $t\geq 0$, where $\lambda>0$ and $\delta>0$ are tuning parameters controlling the amount of regularization and the degree of concavity, respectively.
With a proper choice of $(\lambda,\delta)$, minimizing \eqref{eq:kloss} will force some $\theta_i$'s to be very close to one another, effectively merging these points into a cluster. SPC utilizes a variant of the majorization-minimization (MM) algorithm \cite{deleeuw1994, lange2004}, where the penalty term in \eqref{eq:kloss} is first majorized by a linear function and then the obtained surrogate majorizing function is minimized by cyclic coordinate descent.  

SPC is initialized assuming that all data points form singleton clusters. It then gradually builds up the sparsity in $\| \theta_i - \theta_j \|_2$ and merges the cluster centers by penalizing increasingly larger distances between them. As an output, SPC provides a path of clustering solutions with a decreasing number of clusters, each solution consisting of cluster assignments for all $y_i$ and a number of clusters. Index the solutions on a path of size $S$ by $s = 1, \ldots, S$ and denote the cluster assignment by $A_s=A_s(Y)$ 
and the number of clusters by $\hat{K}^{s}$. The number of total clusters $\hat{K}^{s}$ consists of clusters of any size, including singleton clusters, and $\hat{K}^{s} \geq \hat{K}^{s+1}$ for all $s$. The solution path $\calA(Y) = \{ A_1, \ldots, A_S \}$ is generated using an adaptive data-driven approach by automatically selecting a combination of the penalty parameters $(\lambda, \delta)_s$ for each solution. The choice of the combination of $(\lambda, \delta)_s$ is based on the properties of the MCP and is guided by the pairwise distances between center parameters $\theta_i$'s. Please refer to \cite{marchetti2014} for a detailed description and analysis of SPC.  

The noisy data in each solution $A_s$ can be identified as singleton or very small clusters, i.e. clusters of size $N_k \leq n_0$. Our default choice is $n_0=3$; however, a much larger cutoff cluster size can be selected based on some knowledge about the data. Another special characteristic of SPC is that it can find tight clusters in the initial solutions, similar to hierarchical clustering, when a clustering tree is cut at a smaller distance or at a relatively large number of clusters. Finally, SPC is very easy to use as it effectively has only one tuning parameter that determines the approximate proportion of nearest neighbors to be merged for the initial solution. These characteristics of SPC coupled with the small solution path allow us to develop an iterative subsampling algorithm that can handle large noisy data and can produce a single clustering solution.

\subsection{Sequential cluster assignment}
\label{sec:sequential}

In this section we describe the sequential cluster assignment procedure that can be applied to noisy data and that is thereafter combined with SPC for the full  ISSPC algorithm. The sequential cluster assignment can be directly connected to  classification, specifically to discriminant analysis as it is based on evaluating the likelihood ratios for new data points to determine their cluster memberships. 
We assume that a dataset can generally be separated into data points showing a grouping pattern and data points without any grouping pattern, i.e. noise. We introduce a background model $M_0$ under which a noise data point follows $\mathcal{N}_p( \mu_0, \Sigma_0 )$, while clustered data points are modeled by a mixture Gaussian distribution $M_{\mathcal{C}}$. In short, we apply SPC on a subsample to estimate parameters for these models and then classify the
remaining data points to obtain their cluster memberships. In this sense, the role of the subsample is similar to that of training data and the remaining data similar to test data. So we may also call the two subsets training and test data, respectively.

Suppose that we have drawn without replacement a subsample $\mathscr{D}$ (training data) from $Y$ and denote the remaining data
by $\mathscr{T}$ (test data). Let $D, T \subset \{ 1, \ldots, n\}$ be the indices for data points in $\mathscr{D}$ and $\mathscr{T}$. The SPC method is applied to $\mathscr{D}$ to obtain a clustering path $\calA({\mathscr{D}})$ from which a cluster assignment is selected. Write the selected clustering assignment as $C^{t} = \{ \mathcal{C}, C_0 \}$, where $\mathcal{C} = \{ C_1, \ldots, C_K\}$ denotes the clusters and $C_0 $ contains the indices of the data points identified as noise. Assuming that $y_i$ follows a mixture Gaussian distribution $M_{\mathcal{C}}$, its likelihood is
\begin{align}
\label{eq:Ls}
 &L \left( y_i | M_{\mathcal{C}} \right) \\
 = & \sum_{k=1}^{K} \frac{ \pi_k } { \left| 2\pi\Sigma_k \right|^{1/2} } 
\exp \left\{ -\frac{1}{2} \left( y_i - \mu_k \right)^T \Sigma_k^{-1} \left( y_i - \mu_k \right)  \right\} \nonumber\\
 \defi &\sum_{k=1}^K L_k(y_i | M_k), \nonumber
\end{align}
with $\mu_k$ denoting the mean, $\Sigma_k = \text{diag}( \sigma^2_{k1}, \ldots, \sigma^2_{kp})$ -- a diagonal covariance matrix, and the mixture proportions $\sum _{k=1}^{K} \pi_k = 1$.  Let $N_k=|C_k|$ be the size of a cluster. Based on the clustering assignment $\mathcal{C}$ of the training data, we estimate $\mu_k$ by the cluster sample mean $\bar{y}_k = \frac{1}{N_k} \sum_{i \in C_k} y_i$, $\Sigma_k$ by cluster sample variances $S_{k} = \text{diag}( s^2_{k1}, \ldots, s^2_{kp})$, where  
\begin{equation}\label{eq:svark}
s^2_{km}=\frac{1}{N_k-1} \sum_{i \in C_k} ( y_{im} - \bar{y}_{km} )^2, 
\end{equation}
and the mixture proportions $\hat{\pi}_k = N_k/ \sum_{j=1}^K N_j$.
On the other hand, the background model $M_0$ is estimated by the overall mean and variance of all the data points $Y$, i.e., $\hat{\mu}_0=\bar{y} = \frac{1}{n} \sum_i {y_i}$ and $\hat{\Sigma}_0=S_0 = \text{diag}( s^2_{01}, \ldots, s^2_{0p} )$ with 
\begin{equation}\label{eq:svar0}
s^2_{0m} = \frac{1}{n-1} \sum_{i=1}^n ( y_{im} - \hat{\mu}_{0m} )^2. 
\end{equation}
To simplify notation, we use $\hat{M}_0$ and $\hat{M}_{\mathcal{C}}$ ($\hat{M}_{k}$) to denote
the two models with the estimated parameters.

We then classify the test data $\mathscr{T}$ sequentially based on the estimated models $\hat{M}_0$ and $\hat{M}_{\mathcal{C}}$. We follow a simple decision rule that assigns a test data point $y_i$, $i \in T$, to the more likely model. 
The likelihood ratio for each $y_{i}$, $i \in T$, is
\begin{align}
\label{eq:LRT}
& \Lambda(y_i) = \frac{ L ( y_i | \hat{M}_{\mathcal{C}} ) }{ L ( y_i | \hat{M}_{0} ) } = \frac{ \sum_{k=1}^K L_k(y_i | \hat{M}_k) }{ L ( y_i | \hat{M}_{0} ) }. 
\end{align}
Let $G(y_i)\in\{0,\ldots,K\}$ be the cluster membership for $y_i$. Then the assignment rule is
\begin{align}
\label{eq:DR}
& G(y_i) = \left\{
\begin{array}{cc}
\displaystyle\argmax_{1\leq k\leq K} L_k( y_i | \hat{M}_k ) & \text{if} \quad \Lambda(y_i) \geq c \\
0 & \text{if} \quad \Lambda(y_i) < c
\end{array} \right.,
\end{align}
where by default $c = 1$. The threshold $c$ can also take other values depending on the application; however, we assume $c=1$ for all the simulation and real data studies in this paper. 
Note that $y_i$ is identified as noise if $G(y_i)=0$. 
Once $y_i$ is assigned to a cluster $C_k$, the estimated parameters $\hat{\pi}_k$, $\bar{y}_k$ and $S_k$ are updated for the calculation of \eqref{eq:LRT} for the next test data point. 

The above assignment rule can be regarded as a mixture discriminant analysis. Mixture discriminant analysis \cite{hastie1996} is a generalization of linear and quadratic discriminant analysis, which was further generalized and extended to the models with varying covariance matrices \cite{fraley2002}. In our case, we have a mixture discriminant model with two classes, where one of the classes, the cluster model $M_{\mathcal{C}}$, is a Gaussian mixture model with $K$ components.

\subsection{Iterative subsampling}
\label{sec:subsampleSPC}

We now show how sequential cluster assignment can be combined with SPC to produce accurate clustering solutions for large datasets that otherwise would be computationally expensive or prohibitive. Given a large dataset $Y$, we choose the subsample size $\nu = |\mathscr{D}|= a \sqrt{n}$ where $a \geq 1$ is a small scalar, so that the computational complexity
of both the clustering and the sequential assignment steps is $O(n)$. However, such a small subsample will most probably not be able to capture all the clustering structure of the full data $Y$. Thus, we introduce a recursion between the clustering and the sequential assignment steps. Let $Y_0$ be the data points identified as noise by either the clustering or the sequential steps after the current iteration. That is, the index set of data points in $Y_0$ is
$C_0 \cup \{i \in T: G(y_i)=0\}$.
SPC is then repeated for a random sample of the same size $\nu$ taken from $Y_0$, followed by another sequential 
assignment step. This recursion will be repeated  until no more clusters are found.
Since SPC does not require the number of clusters as an input parameter and provides a short solution path, it is 
essential to select one clustering solution for the sequential step in the combined algorithm. Instead of relying on a 
sophisticated model selection procedure, we consider any solution on the path as
a set of potential clusters $\mathcal{C}$. We test whether these potential clusters are significantly different from the null model $M_0$, which is characterized by a large variance since it is estimated with all the data points, including noise. If some clusters in $\mathcal{C}$ are not significantly different from the null, then the data points in these clusters are reassigned to $C_0$. This quality check becomes important for later stages in the clustering-assignment recursion since the clustering is assumed to be performed on an increasing amount of noise. It also becomes a stopping criterion for the recursion: the recursion is terminated when no more clusters are found to be significantly different from the null. 

Now we describe the detailed testing procedure. For a particular solution, we compare each estimated covariance matrix $S_k$, $k=1,\ldots,K$, to the estimated null covariance matrix $S_0$. Since we have assumed that both $\Sigma_k$ and $\Sigma_0$ are diagonal, a set of $p$ hypothesis tests can be performed for each cluster $k$:
\begin{align*}
H^0_{km}: \sigma^2_{km} = \sigma^2_{0m} \quad \text{vs} \quad H^1_{km}:  \sigma^2_{km} < \sigma^2_{0m},
\;m=1,\ldots,p, 
\end{align*}
where we reject $H^0_{km}$ if $s^2_{km}$ \eqref{eq:svark} is sufficiently small relative to $s^2_{0m}$ \eqref{eq:svar0}. Note that $s^2_{0m}$ and $s^2_{km}$ are computed from the same data and therefore are not independent. However, since by design $n \gg \nu>N_k$ for all $k$, the variance of $s^2_{0m}$ is negligible compared to $s^2_{km}$ and thus may be treated as a constant. Then the test statistic
\begin{align}
\label{eq:ftest}
& F_{km} = \frac{(N_k-1) s^2_{km}}{s^2_{0m}}, \;\;k=1,\ldots,K \text{ and } m=1,\ldots,p,
\end{align}
follows approximately a $\chi^2$-distribution with $N_k - 1$ degrees of freedom if $H^0_{km}$ is true. Next, we sort the p-values $P_{km}$ of the statistic $F_{km}$ in ascending order, $P_{k(m)} \leq P_{k(m+1)}$, and apply the Benjamini-Hochberg procedure to find
\begin{align}
\label{eq:fdr}
m^*_k = \max \left\{ m : P_{k(m)} \leq \frac{m}{p} \beta, m=1,\ldots,p \right\}
\end{align} 
for each $k$, where $\beta$ is the cutoff of the false discovery rate (FDR). We keep cluster $C_k$ if $m^*_k \geq \eta$ with $\eta \in (1,p)$, i.e. if we find sufficiently many dimensions with significantly smaller $s^2_{km}$ compared to the corresponding $s^2_{0m}$. Otherwise, we discard cluster $C_k$ and reassign its members to $C_0$. The procedure of controlling the FDR for each cluster allows us to account for the relevancy of a subset of dimensions or features in each cluster, which becomes more critical for high-dimensional data. In addition, this procedure will generally reject very small clusters with moderately small variances, and thus, it can also be seen as a way of controlling for a minimum cluster size. Finally, if all the clusters are discarded after the SPC step for a subsample, then the clustering-assignment recursion is terminated.

The hypothesis testing coupled with FDR control effectively becomes a practical mechanism for an automatic determination of the number of clusters in a dataset. In fact, we do not need to use any sophisticated solution selection methodology in order to pick a solution from $\calA(\mathscr{D})$, but we would simply select the first solution with the largest number of clusters of size $N_k >n_0$, followed by the above testing procedure to remove loose clusters. This gives us a higher chance of discovering tight clusters in the presence of noise. The effectiveness of this approach will be demonstrated in Sections~\ref{sec:simstudies} and \ref{sec:genedata}. 

The full ISSPC method is outlined in Algorithm~\ref{alg:isspc}, in which $b$ is the iteration number in the clustering-assignment recursion and $\omega\in(0,1)$ is an input parameter for the SPC algorithm. The collection of clusters generated in the $b$th iteration is denoted by $\mathcal{C}^{(b)}$, while the noise data points are denoted by $Y_0^{(b)}$. The set $\mathcal{C}$ consists of all the clusters found along the recursion.

\algsetup{indent=1.5em}
\begin{algorithm}[t]
\caption{Iterative Subsampling SPC (ISSPC)}
\label{alg:isspc}
\begin{algorithmic}
\STATE \textbf{Inputs}
\STATE{input: $Y = (y_{im})_{n \times p}$, $\omega^{(1)} \in (0,1)$, $\nu$, $\eta$}
\STATE{default input: $\omega^{(b)} = 0.1$ for $b\geq 2$, $\beta =0.01$}
\STATE{initialization: $b=1$, $Y^{(1)}_{0} = Y$, $\mathcal{C} = \emptyset$}
\end{algorithmic}
\bigskip

\begin{algorithmic}[1]
\REPEAT
\STATE{$\mathcal{C}^{(b)} = \emptyset$}
\STATE{draw a random sample $\mathscr{D}^{(b)}$ of size $\nu$ from $Y^{(b)}_0$}
\STATE{$\mathscr{T}^{(b)}=Y^{(b)}_0\setminus \mathscr{D}^{(b)}$}
\STATE{run SPC algorithm to obtain a solution path $\calA(\mathscr{D}^{(b)})$}
\STATE{choose $A_t=\{C_0,C_1,\ldots,C_K\}\in\calA(\mathscr{D}^{(b)})$ such that $K$ is maximized.}
\FOR{ $k = 1, \ldots, K$ }
\STATE{compute $m^*_k$ as in \eqref{eq:fdr}}
\STATE{ if $m^*_k \geq \eta$, $\mathcal{C}^{(b)} \leftarrow \mathcal{C}^{(b)} \cup \{C_{k}\}$}
\STATE{ else $C_0\leftarrow C_0 \cup C_k$}
\ENDFOR
\IF{$K=0$ or $m^*_k < \eta$ for all $k$} 
\BREAK 
\ENDIF
\FORALL{ $y_i \in \mathscr{T}^{(b)}$ }
\STATE{compute $G(y_i)$ as in \eqref{eq:DR}}
\STATE{assign $y_i$ to $C_0$ or the clusters in $\mathcal{C}^{(b)}$ accordingly} 
\ENDFOR
\STATE{$Y_0^{(b+1)}=C_0$, $\mathcal{C}\leftarrow \mathcal{C}\cup \mathcal{C}^{(b)}$,
$b \leftarrow b+1$}
\UNTIL{$|Y_0^{(b)}| < \nu$}
\end{algorithmic}	
\end{algorithm}

\subsection{The choice of the tuning parameters}
\label{sec:tuning}

Along with the tuning parameter $\omega$ of the original SPC, ISSPC has additional user-defined parameters. These include $\beta$, the FDR cutoff, and $\eta$, the number of dimensions with FDR $< \beta$ that determines whether a cluster should be kept or removed. The choice of these parameters is relatively straightforward, especially for $\beta$, which does not have a big impact on the procedure as long as its value is reasonably low, e.g. $\beta \in (0.1, 0.01)$. Throughout the simulated data and real data examples we set $\beta=0.01$, which is commonly used in practice for hypothesis testing. High values of $\beta$ can result in the acceptance of ``false" clusters, i.e. clusters consisting mostly of noise, and the subsequent erroneous assignment of clustered data into these clusters along with noise. The parameter $\eta$ can be set based on user knowledge of dimension relevance, but we generally recommend to choose $\eta \in (0.1p, 0.5p)$. If $\eta \approx p$, many relevant clusters could be omitted, and if $\eta \approx 1$, too many ``false" clusters can be accepted since low variances can occur by chance in just a few dimensions. It is clear that the parameter $\eta$ can also affect the number of clusters, with smaller values resulting in more clusters.  

It is important to point out that SPC itself and the sequential assignment rely on the assumption that the clusters are generated by a mixture Gaussian distribution. If this assumption is violated, the clustering result might be unsatisfactory, especially considering the fact that the subsample sizes for clustering are very small. A typical example of such an adverse scenario is the presence of several outliers in the SPC clusters, which can trigger the sequential assignment step to incorporate noise or misclassify clustered data. To overcome such problems we may calculate \eqref{eq:LRT} by using robust estimates of $\mu_k$ and $\Sigma_k$, such as the trimmed mean and variance. Trimming 5\% to 25\% of data points is usually an acceptable amount. If $\alpha$\% of data points are removed in trimming, then $(\alpha/2)$\% of the smallest and $(\alpha/2)$\% of the largest values in each dimension are discarded. We perform trimming on clusters with a relatively substantial size; in particular, we set this size to be $N_k > 10 n_0$ to avoid trimming very small clusters that can lead to unstable estimates of means and variances. Generally, more clusters with a smaller size will be discovered with increasing amounts of trimming. This is due to the fact that fewer test data points will be added to the existing trimmed clusters, and these omitted data will most probably be clustered in the next iteration of the subsample clustering. Since the same amount of trimming is done across all of the clusters with a larger size, it is possible that, while very tight patterns can be extracted, some larger clusters could be split as a result.          

To conclude, we would like to comment on the choice of the original SPC tuning parameter $\omega$. This parameter can be determined by the user for the first iteration of Algorithm~\ref{alg:isspc}; however, we generally recommend $\omega=0.1$ or a comparably small value for $b\geq 2$. Low values of $\omega$ will ensure that smaller clusters with fewer outliers are discovered in later iterations and the procedure does not terminate prematurely.

\section{Simulation studies}
\label{sec:simstudies}

In this section we demonstrate the performance of ISSPC on simulated data. We generate large datasets and cluster these data with the original SPC algorithm, ISSPC, the mclust method in the R package $\verb+mclust+$ \cite{fraley2002}, and the Scalable Weighted Iterative Flow-clustering Technique (SWIFT) \cite{naim2014}. We chose mclust based on its ability to produce high quality results, to separate noisy data points and to estimate the number of clusters, as well as its competitive speed as shown in \cite{marchetti2014}. We also selected SWIFT, a recently developed algorithm, due to its ability to scale to large multi-dimensional datasets, to isolate small distinct groups, and to estimate the number of clusters in a dataset automatically. It is also conceptually related to ISSPC as it uses a subsampling procedure in combination with the EM algorithm in order to reduce computational complexity. Another benefit of the SWIFT algorithm is that it is readily available as a Matlab implementation. We compare the accuracy and the speed of both SPC approaches, mclust and SWIFT. The accuracy is evaluated based on the adjusted rand index (ARI) developed in \cite{hubert1985}, which compares an estimated cluster assignment to the true cluster assignment. We calculate two different ARI scores $(\text{ARI}_c, \text{ARI}_n)$ as stated in the Appendix to gauge the performance with respect to the clustered data and noise. $\text{ARI}_c$ assesses whether data points are misclassified, whether identified clusters are merged or split, and whether the estimated clusters contain noise. $\text{ARI}_n$ evaluates whether any clustered data point is misclassified as noise. 
    
\begin{figure*}
	\centering
	\begin{subfigure}[b]{0.45\textwidth}
		\centering
		\includegraphics[width=\textwidth]{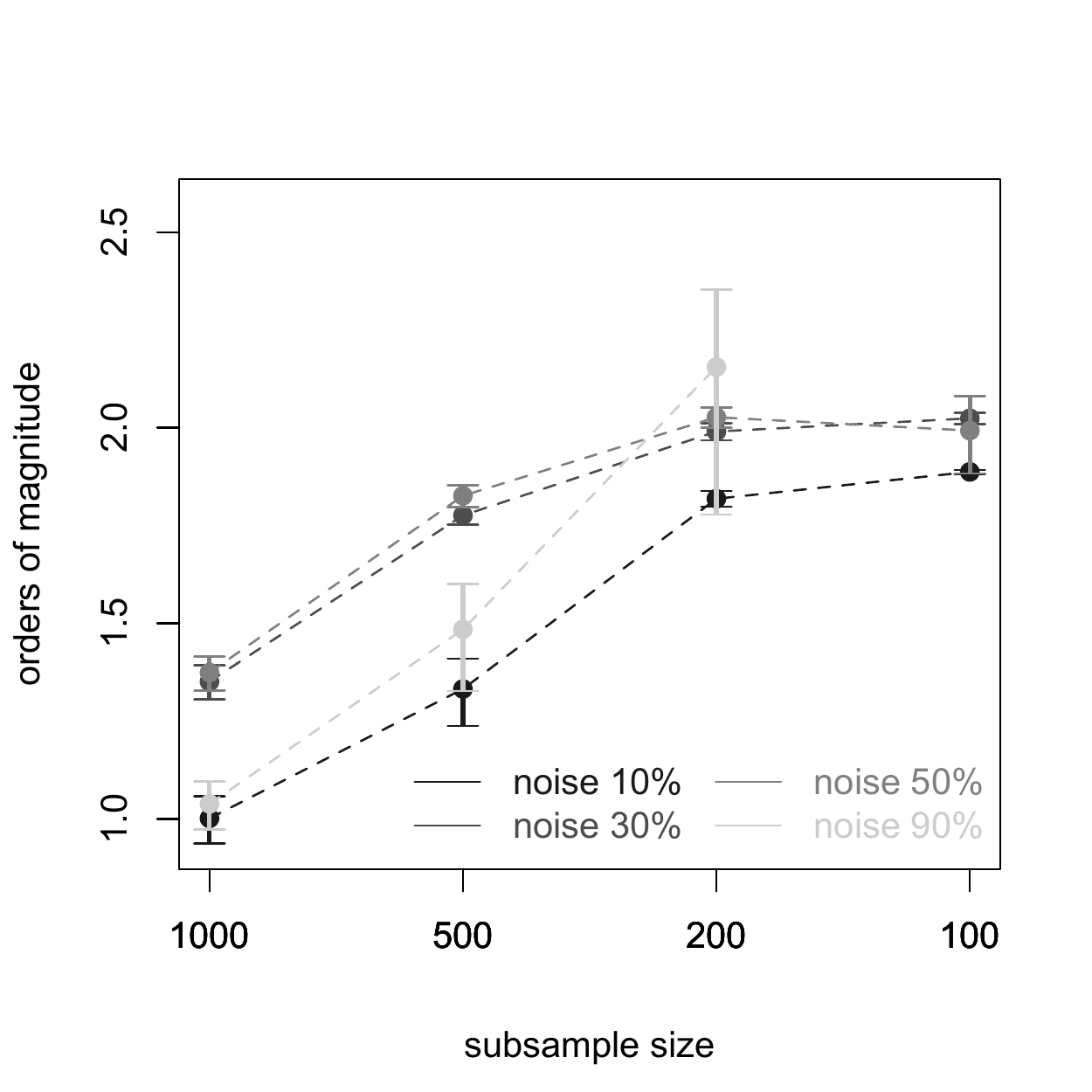}
		\caption{$n=10,000$}
		\label{om10k}
	\end{subfigure} %
	\begin{subfigure}[b]{0.45\textwidth}
		\centering
		\includegraphics[width=\textwidth]{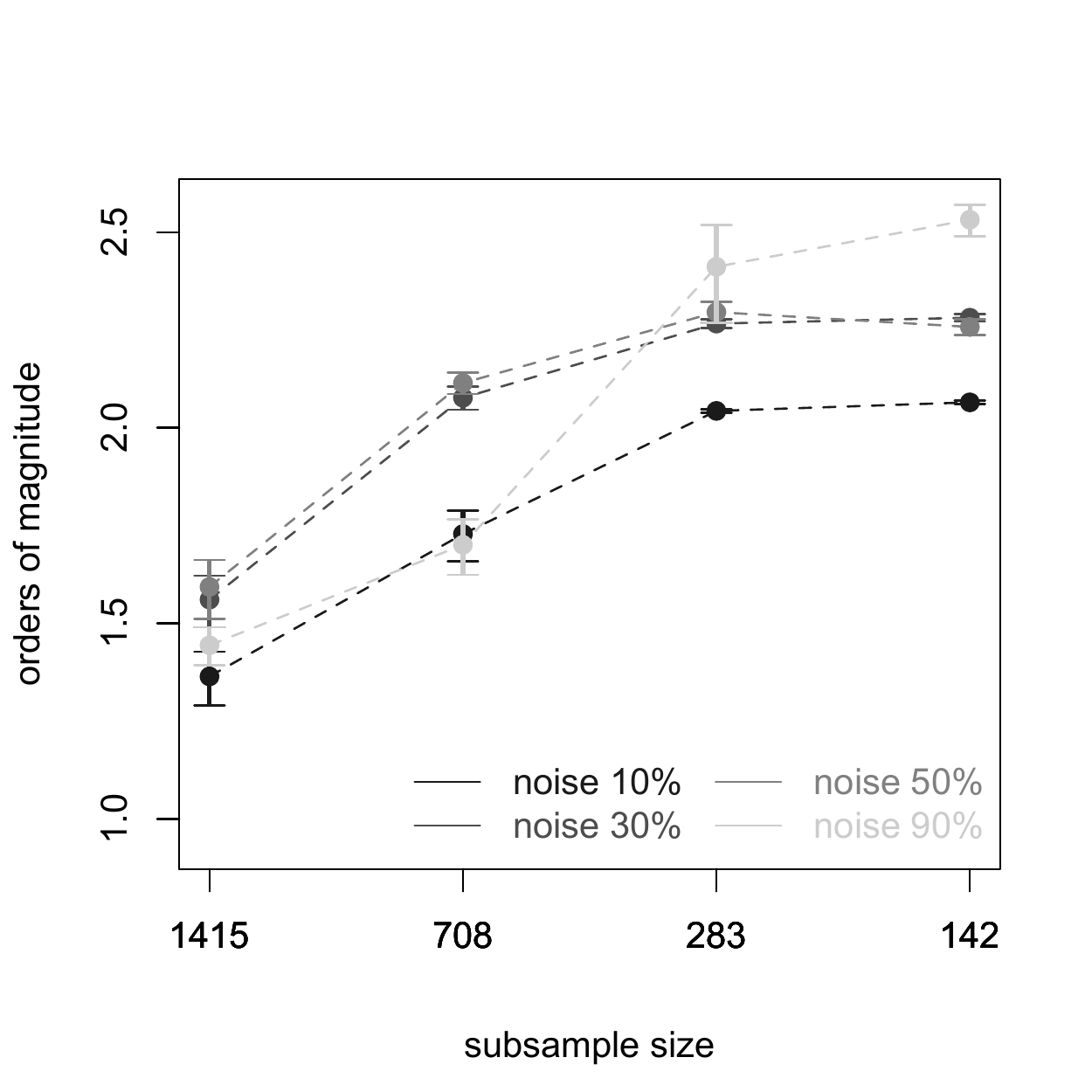}
		\caption{$n=20,000$}
		\label{om20k}
	\end{subfigure} %
	\caption{The orders of magnitude of computational savings of the ISSPC algorithm compared to the original SPC algorithm. The bars indicate 95\% confidence intervals. The subsample sizes are $a\sqrt{n}$ for $a=10,5,2,1$ from
	left to right. }
	\label{om}
\end{figure*}  

We performed the comparison on four different data sizes -- $n=10,000$, $n=20,000$, $n=50,000$, and $n=100,000$ and $p=20$ dimensions, and generated each dataset with four different proportions of noise -- $0.1n$, $0.3n$, $0.5n$, and $0.9n$. The clustered data in each of a total of 16 datasets were generated from a Gaussian mixture model with $K=10$ clusters of about equal size and variance. The noise was generated from a uniform distribution on $[ -5, 5 ]^p$ outside the radiuses of the clusters, where the radius of a cluster is defined as the largest distance from the cluster center to the data points in that cluster. The ISSPC algorithm was run 20 times independently on each dataset, and the ARI scores and the computation speed are reported as an average of the 20 runs, with each run using different subsamples. We chose $\eta =10$ since all the dimensions were relevant for all the simulated clusters. We were not able to apply the original SPC algorithm to the datasets of size $n > 20,000$ due to the operating memory limitations in holding big distance matrices and present the comparison results only with $n=10,000$ and $n=20,000$. The results from $n=50,000$ and $n=100,000$ are presented to demonstrate the accuracy, stability and potential time savings of the ISSPC approach. 

Mclust, when applied to data with noise, has in effect three inputs that include the categorization of data into noise and clusters, a range of the number of clusters for model selection via Bayesian Information Criterion (BIC), and the reciprocal hyper volume $V$ of the data region. For more details on the tuning parameters of mclust, please refer to \cite{fraley2002}. The recommended $K$th nearest neighbor cleaning method was applied for the categorization \cite{byers1998} with $K=10$ and the range of the number of clusters was set to $(5,15)$. We could not use the default value for $V$, which appeared to be too small and forced all the noise into the clusters. Instead, we estimated a larger $V$ using the approximate volume of the convex hull of the data, which was computed with the $\verb+geometry+$ package in R based on six dimensions and extrapolated roughly to 20 dimensions.     

Lastly, SWIFT consists of three stages: weighted iterative sampling based EM, multimodality splitting and agglomerative merging. In the first stage, the EM algorithm is applied repeatedly to different subsamples of data, drawn based on weights derived from the posterior probability of each estimated cluster. Large clusters are then ``fixed", and the subsampling procedure is applied to the remaining data points, where preference is given to those data points unlikely to come from any fixed clusters. Once all the clusters become fixed, a few iterations of the EM are applied to the full dataset. In the second splitting stage, the algorithm separates a multimodal cluster into multiple unimodal clusters, and in the third stage it merges some clusters from the previous stage based on linear discriminant analysis. This last stage is aimed at detecting clusters with non-convex shapes. It must also be noted that even though SWIFT does not isolate noise in the data in a direct way, it can identify so-called ``background" clusters, which are defined as having large volume and low density. In other words, SWIFT partitions all the data, however, some clusters might consist entirely or almost entirely of the noisy data points. These ``background" clusters could be equated to noise for our scenarios for the accuracy comparisons. SWIFT requires very little fine tuning, and we only needed to specify the initial estimate for the number of clusters $K_0$. In fact, we used the default $K_0 = 100$ since smaller values resulted in slower computation times in most scenarios with minor gains in accuracy. For example, a simulated dataset of size $n=10,000$ and noise of $0.5n$ took about $23$ minutes to process with $K_0=50$ versus $3$ minutes with $K_0=100$. The authors mention that the splitting stage is the most computationally intensive step, and smaller values for $K_0$ result in a larger number of splits, which might be the reason for the decrease in speed. 
SWIFT provided two separate solutions from its merging and splitting stages, correspondingly. We used the solution from the merging stage for comparison as it is the final step of the algorithm.

\subsection{Computation time}
\label{sec:time}

First, we show computational savings of the new approach compared to the original algorithm for $n=10,000$ in Figure~\ref{om10k} and for $n=20,000$ in Figure~\ref{om20k}. The curves show the average orders of magnitude of computational savings compared to the original SPC algorithm. The different gray scales in the plots represent the different proportions of noise in the datasets. It can be seen from Figure~\ref{om} that computational improvements of this procedure are considerable and can amount to somewhere between 1 to 2 orders of magnitude (10 to 100 times faster) for $n=10,000$ and even larger, 1.5 to 2.5 orders of magnitude, for $n=20,000$.  
For the dataset with 90\% of noise and of size $n=10,000$, ISSPC with a subsample size of $\sqrt{n}=100$ did not find any clusters, and thus the computation time for this case is not reported in Figure~\ref{om10k}.

\begin{figure*}
	\centering
	\begin{subfigure}[b]{0.4\textwidth}
		\centering
		\includegraphics[width=\textwidth]{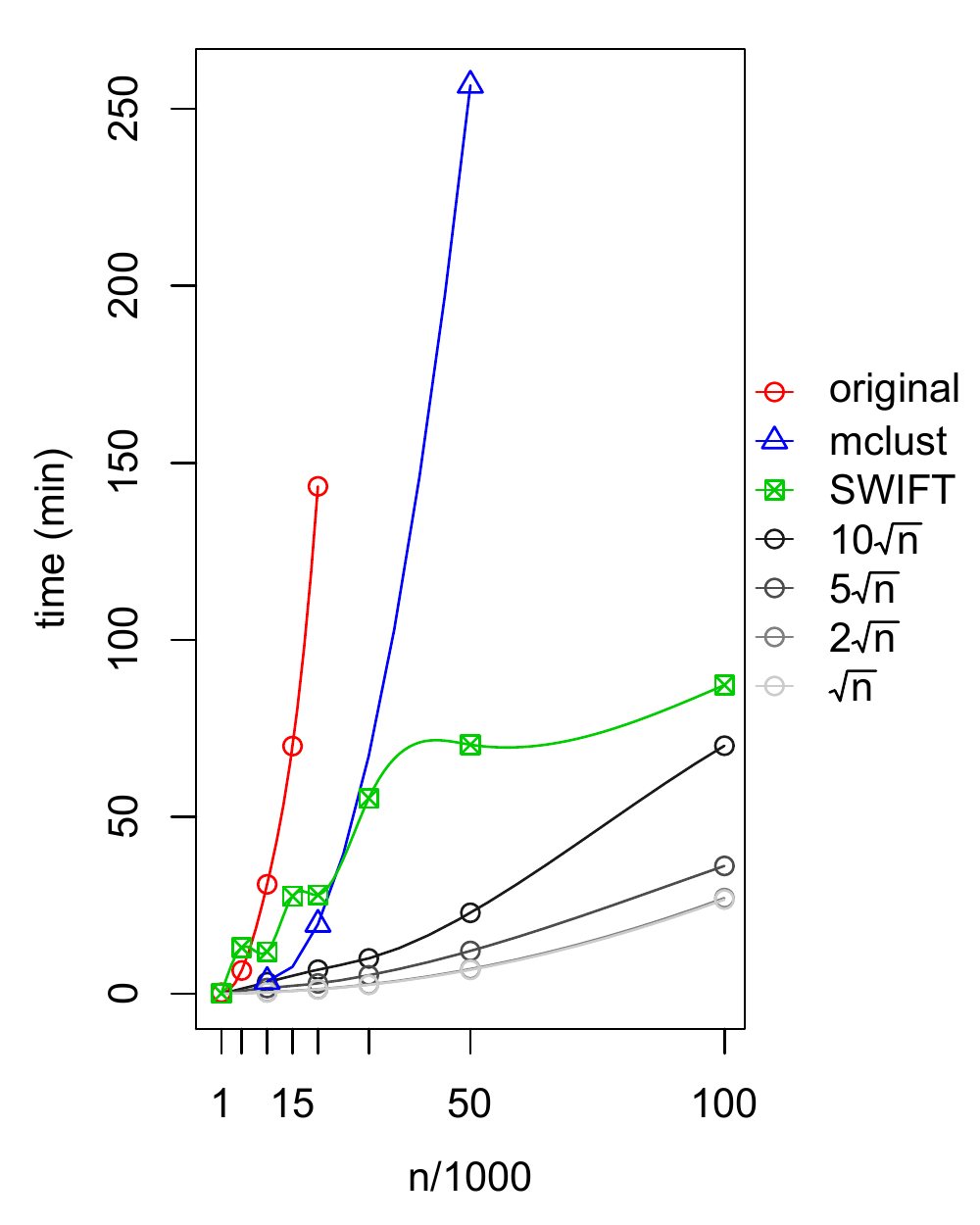}
		\caption{noise $0.1n$}
		\label{runtimes_01}
	\end{subfigure} %
	\begin{subfigure}[b]{0.4\textwidth}
		\centering
		\includegraphics[width=\textwidth]{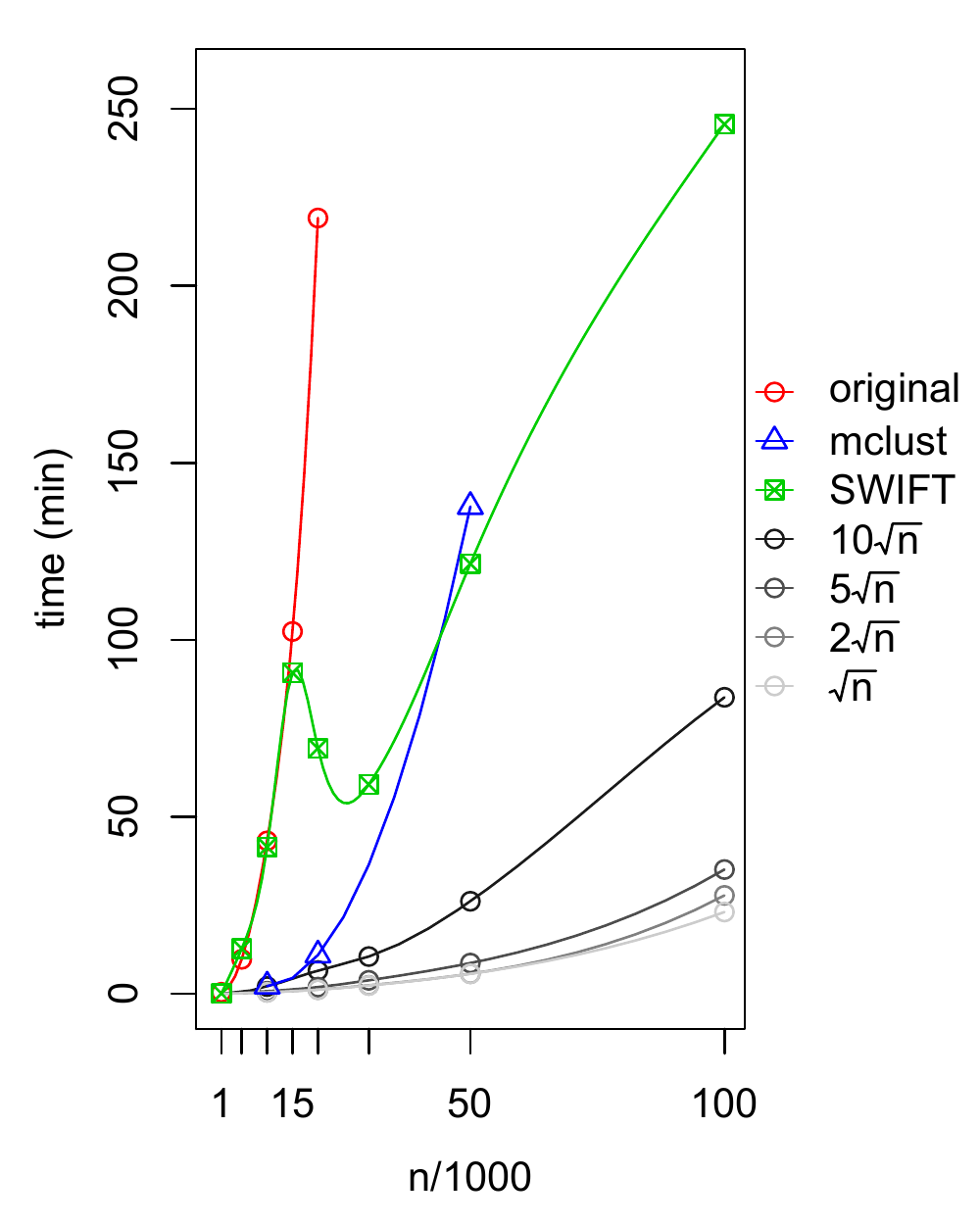}
		\caption{noise $0.3n$}
		\label{runtimes_01}
	\end{subfigure} %
	\begin{subfigure}[b]{0.4\textwidth}
		\centering
		\includegraphics[width=\textwidth]{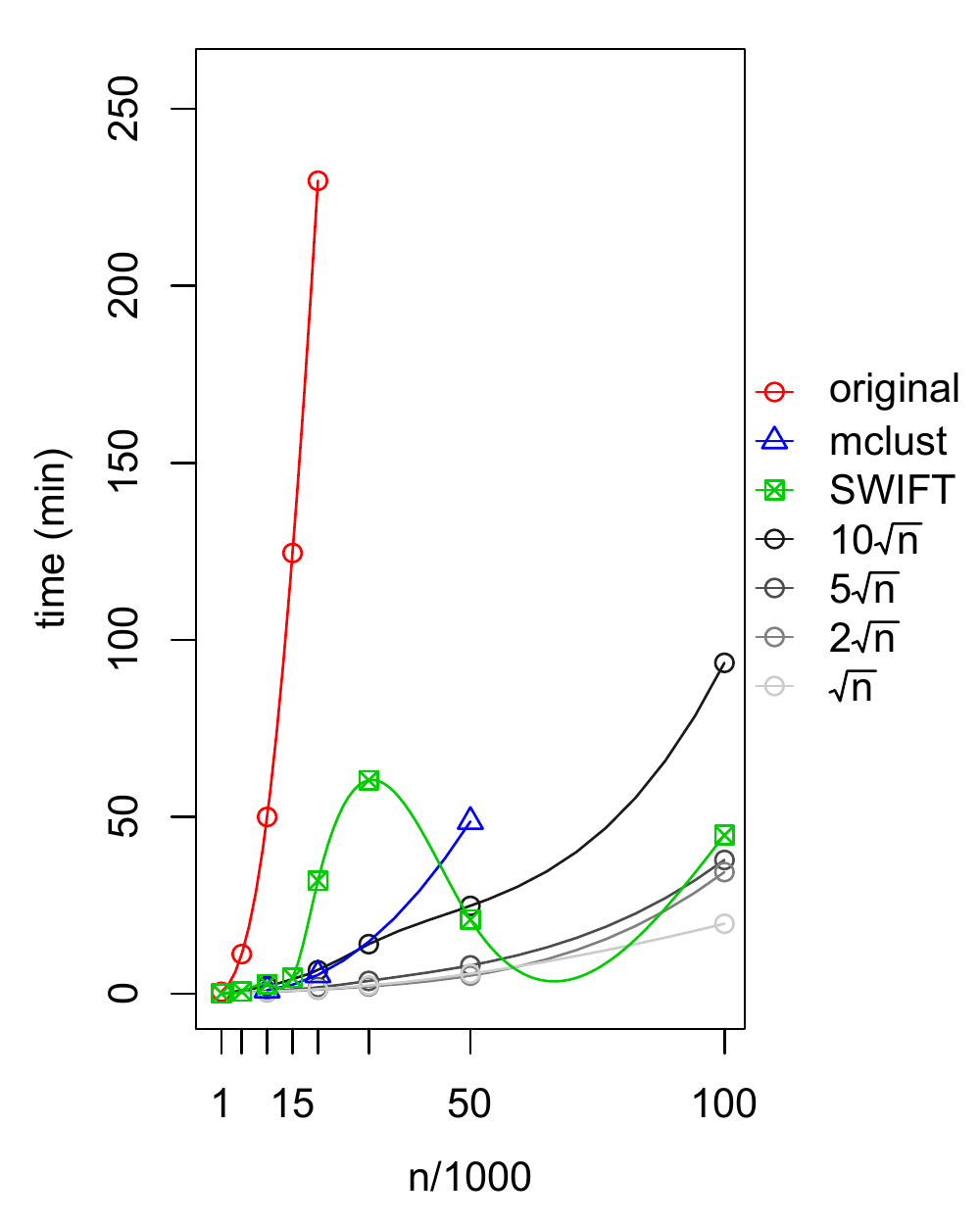}
		\caption{noise $0.5n$}
		\label{runtimes_01}
	\end{subfigure} %
	\begin{subfigure}[b]{0.4\textwidth}
		\centering
		\includegraphics[width=\textwidth]{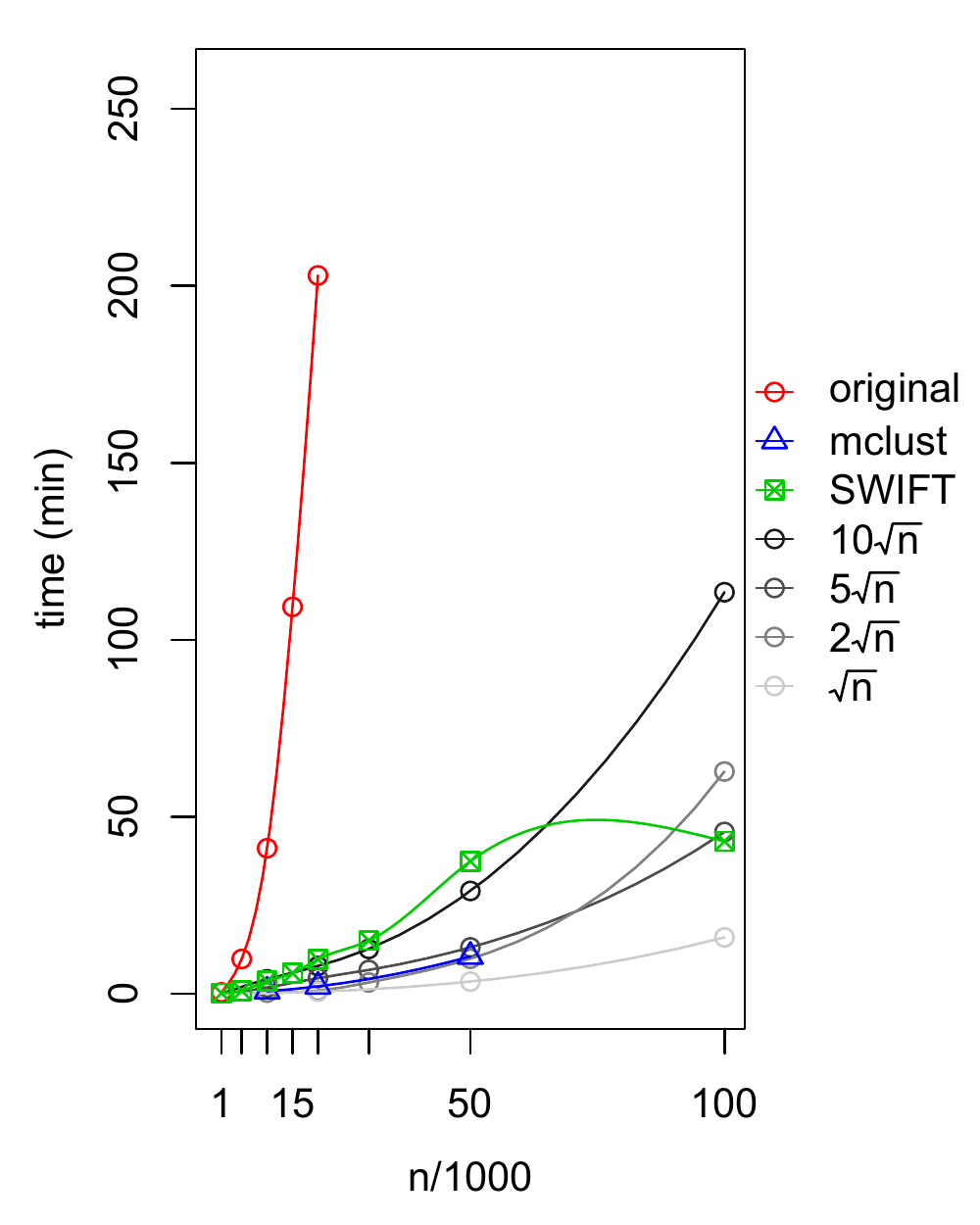}
		\caption{noise $0.9n$}
		\label{runtimes_01}
	\end{subfigure} %
	\caption{The run times in minutes for the original SPC (red), mclust (blue), SWIFT (green) and ISSPC (gray scale). Each panel corresponds to a different noise proportion in the datasets.}
	\label{runtimes}
\end{figure*}

Figure~\ref{runtimes} shows the comparison of the absolute run times for all the compared clustering methods across all the simulated datasets. The run time for mclust includes the $K$th nearest neighbor procedure that estimates the noisy data points. It can be seen from the figure that the run time of mclust depends on the amount of noise and is much shorter when this amount is large due to the fact that mclust is effectively applied only to the identified clustered data. It should be noted that mclust could not be run on $n=100,000$, due to reaching operating memory limits. The computation times of SWIFT fluctuate, which is probably due to the subsampling procedure, nonetheless, it is clear that it has competitive speed for large datasets. SWIFT repeatedly assigned all the data points to a single cluster with $0.5n$ and $0.9n$ noise for $n=100,000$, however, we still report the computation times for these scenarios. The original SPC algorithm is shown to have a computational time 
complexity of $O(n^2)$ approximately for all the simulated data scenarios with a slightly shorter time when there is little noise. Mclust unsurprisingly shows much better speed than the original SPC algorithm. ISSPC, however, outperforms mclust on speed, especially with larger data sizes. ISSPC is able to cluster a large dataset, e.g. $n=100,000$, in a reasonable amount of time, even when the subsample size is $\nu = 10 \sqrt{n}$, and it can favorably compete with mclust and SWIFT in terms of speed, especially considering the fact that ISSPC is currently implemented in R, while mclust is implemented in Fortran and SWIFT utilizes Matlab's parallel computing toolbox.

\subsection{Accuracy}
\label{sec:accuracy}

\begin{figure*}
	\centering
	\begin{subfigure}[b]{0.45\textwidth}
		\centering
		\includegraphics[width=\textwidth]{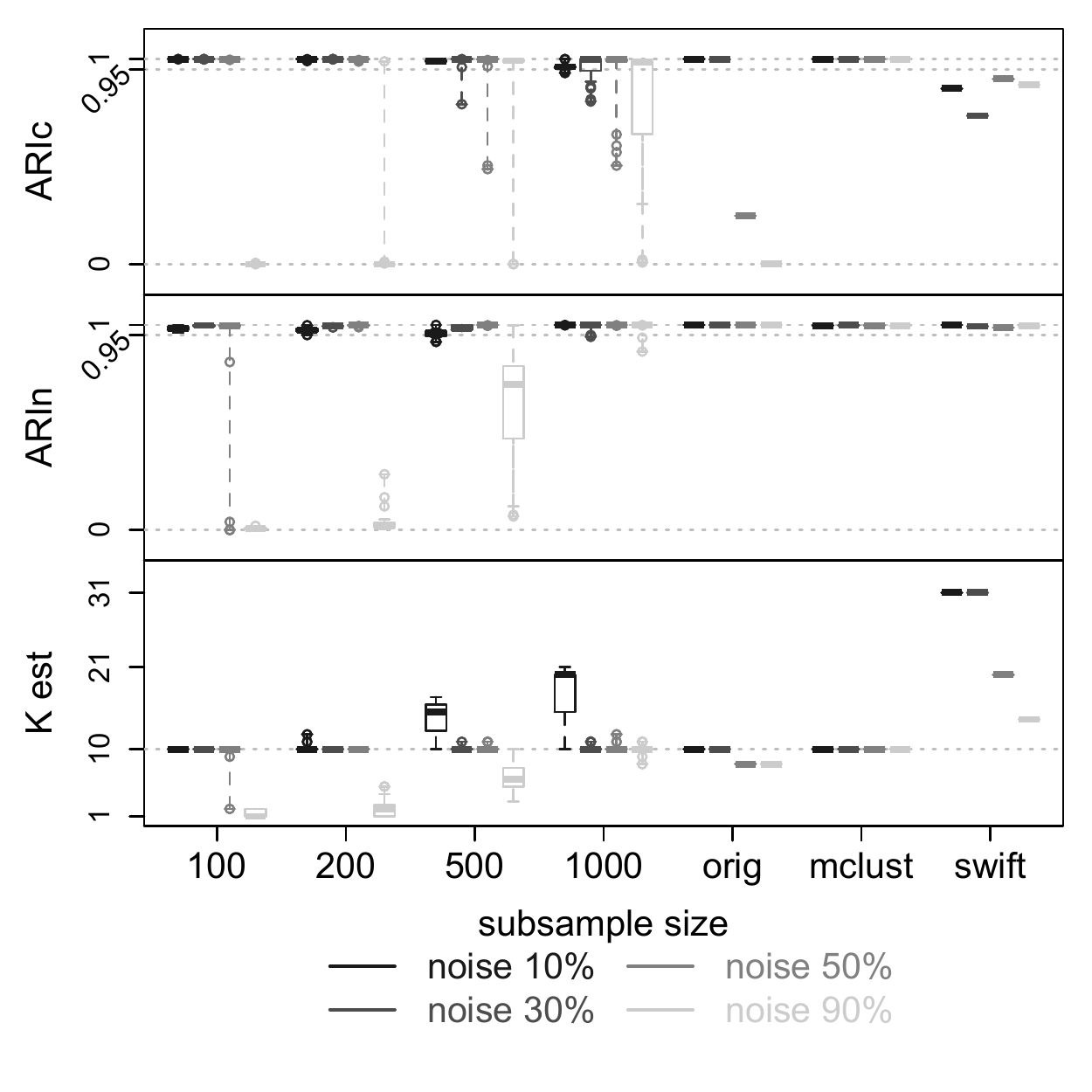}
		\caption{$n=10,000$}
		\label{ari10k}
	\end{subfigure} %
	\begin{subfigure}[b]{0.45\textwidth}
		\centering
		\includegraphics[width=\textwidth]{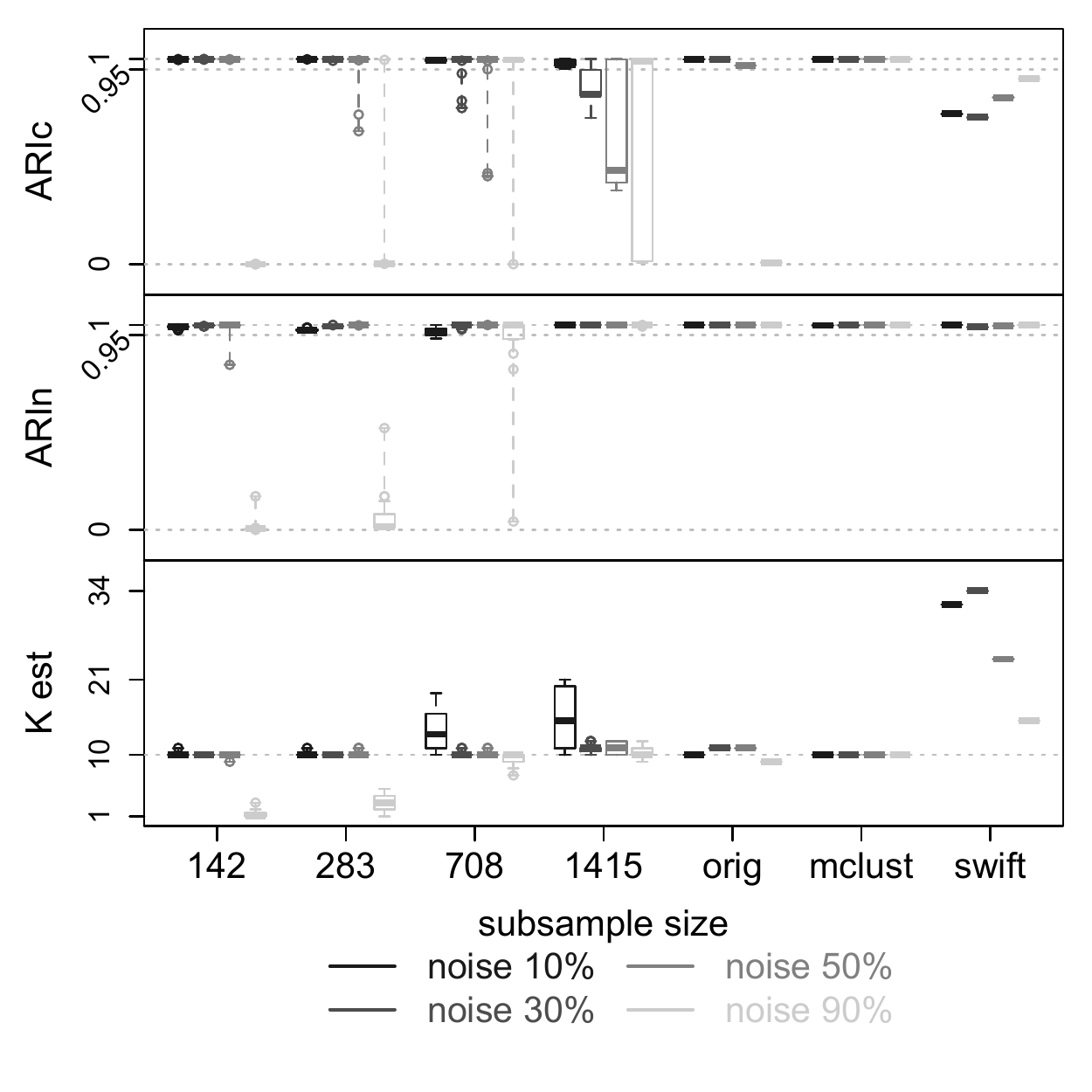}
		\caption{$n=20,000$}
		\label{ari20k}
	\end{subfigure} %
	\begin{subfigure}[b]{0.45\textwidth}
		\centering
		\includegraphics[width=\textwidth]{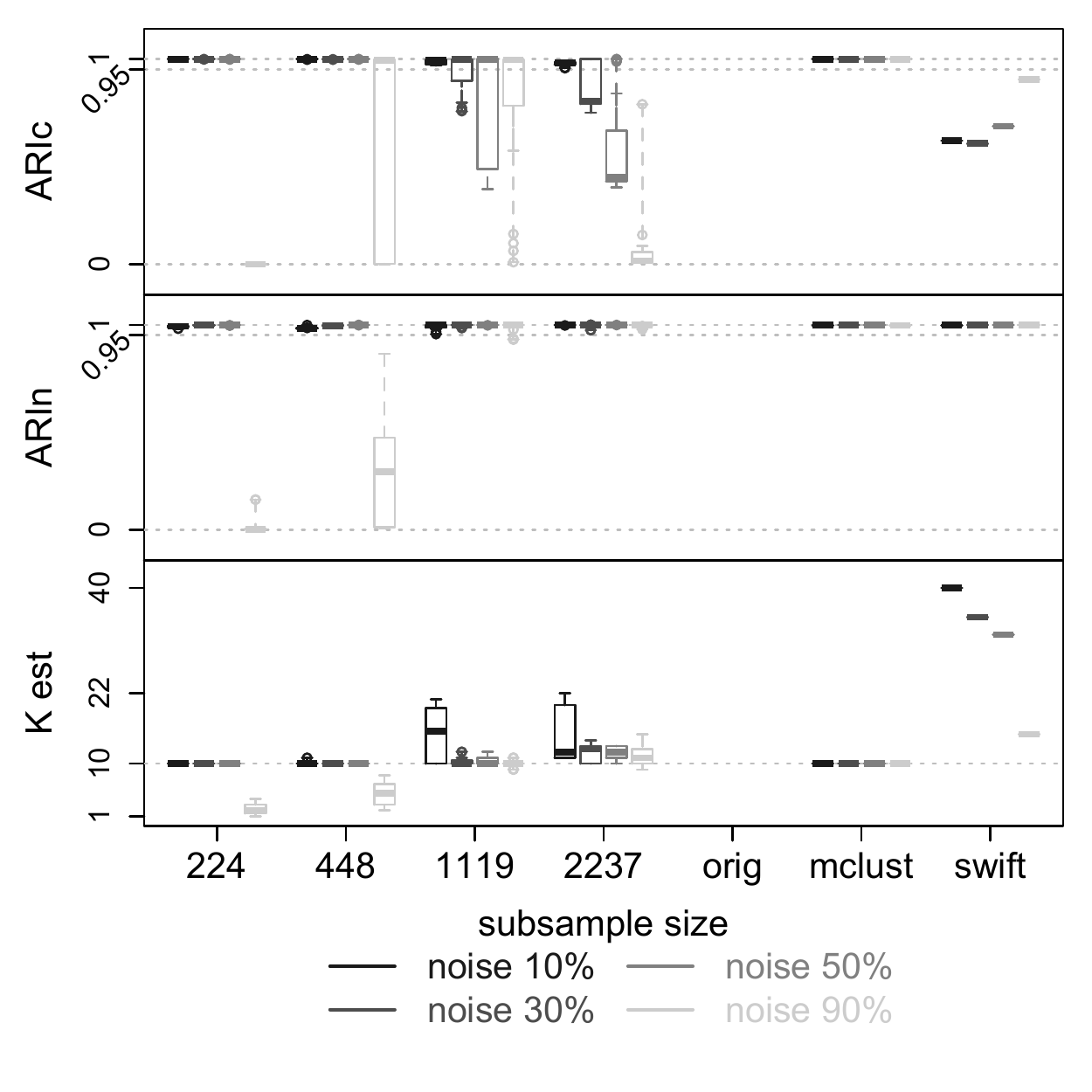}
		\caption{$n=50,000$}
		\label{ari50k}
	\end{subfigure} %
	\begin{subfigure}[b]{0.45\textwidth}
		\centering
		\includegraphics[width=\textwidth]{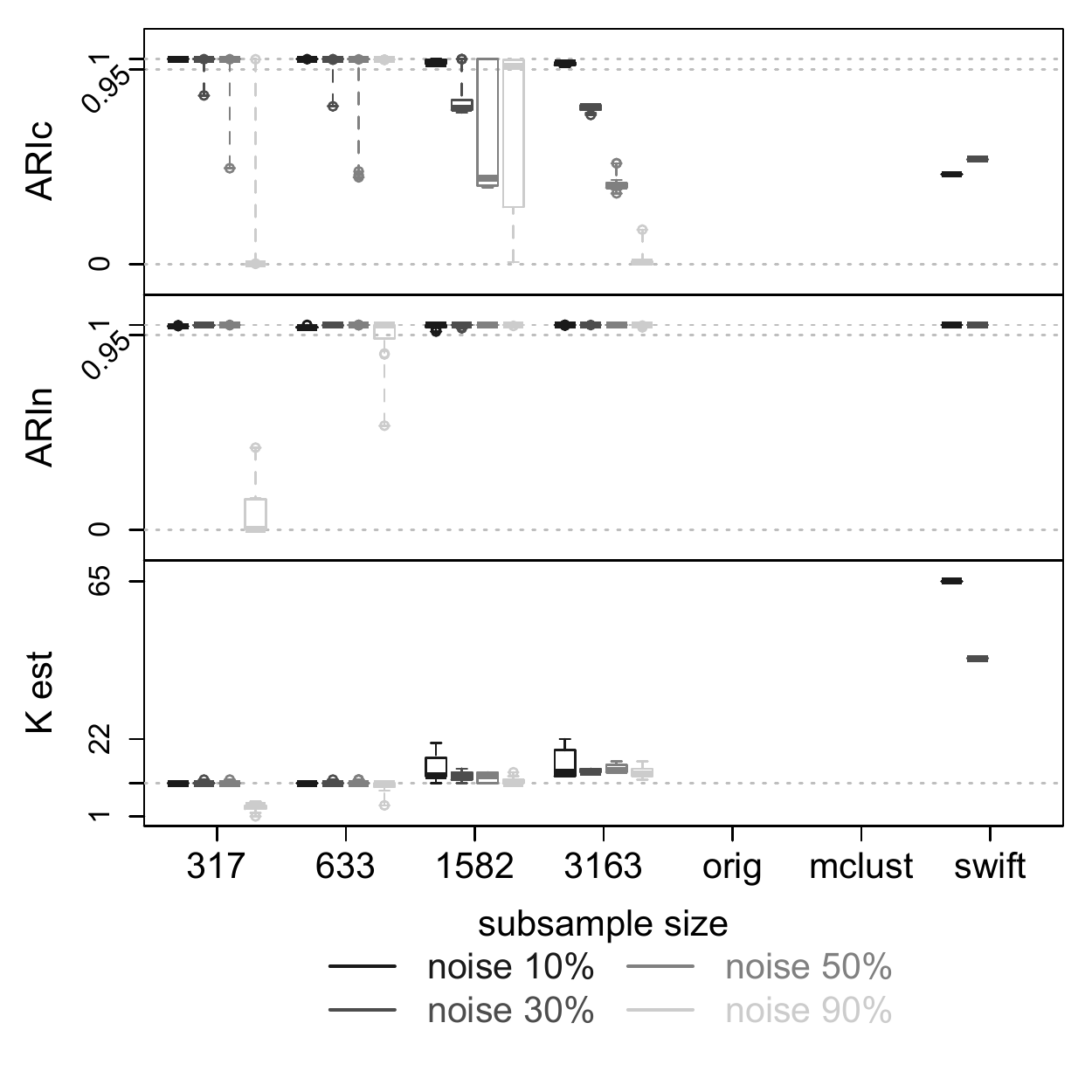}
		\caption{$n=100,000$}
		\label{ari100k}
	\end{subfigure} %
	\caption{ARI scores and the number of estimated clusters for ISSPC, original SPC, mclust and SWIFT for various dataset sizes. From left to right the subsample
	sizes are $\sqrt{n}$, $2\sqrt{n}$, $5\sqrt{n}$, and $10\sqrt{n}$.}
	\label{ari}
\end{figure*}

We now report the comparative performance of original SPC, ISSPC, mclust and SWIFT. Figures~\ref{ari10k}-\ref{ari100k} demonstrate $\text{ARI}_c$ (top panel), $\text{ARI}_n$ (middle panel) and the estimated number of clusters (bottom panel) for each data size $n$. The boxplots for ISSPC show the results across 20 independent runs on the same dataset, while only one (deterministic) result is reported for each dataset for the other methods. The color in the gray scale represents the amount of noise in the dataset, the lighter indicating a higher proportion of noise. The results for the original SPC algorithm are reported after the hypothesis testing and cluster selection procedure, described in Section~\ref{sec:subsampleSPC}, is applied to the solution with the largest number of clusters. SWIFT assigns all the data points into clusters, including ``background" clusters. The ``background" clusters would then ideally contain all the noisy data points. In order to compute $\text{ARI}_n$ scores for SWIFT, we identified clusters with 5\% or fewer clustered data points based on the true assignment as ``background" clusters and re-classified all the data points in those clusters as noise. The estimated number of clusters for SWIFT is based only on the true clusters, with the ``background" clusters excluded. We also note that the $\text{ARI}$ scores are not reported for SWIFT for $0.5n$ and $0.9n$ with $n=100,000$ since all the data points were assigned to a single cluster.

\begin{figure*}
	\centering
	\begin{subfigure}[b]{0.45\textwidth}
		\centering
		\includegraphics[width=\textwidth]{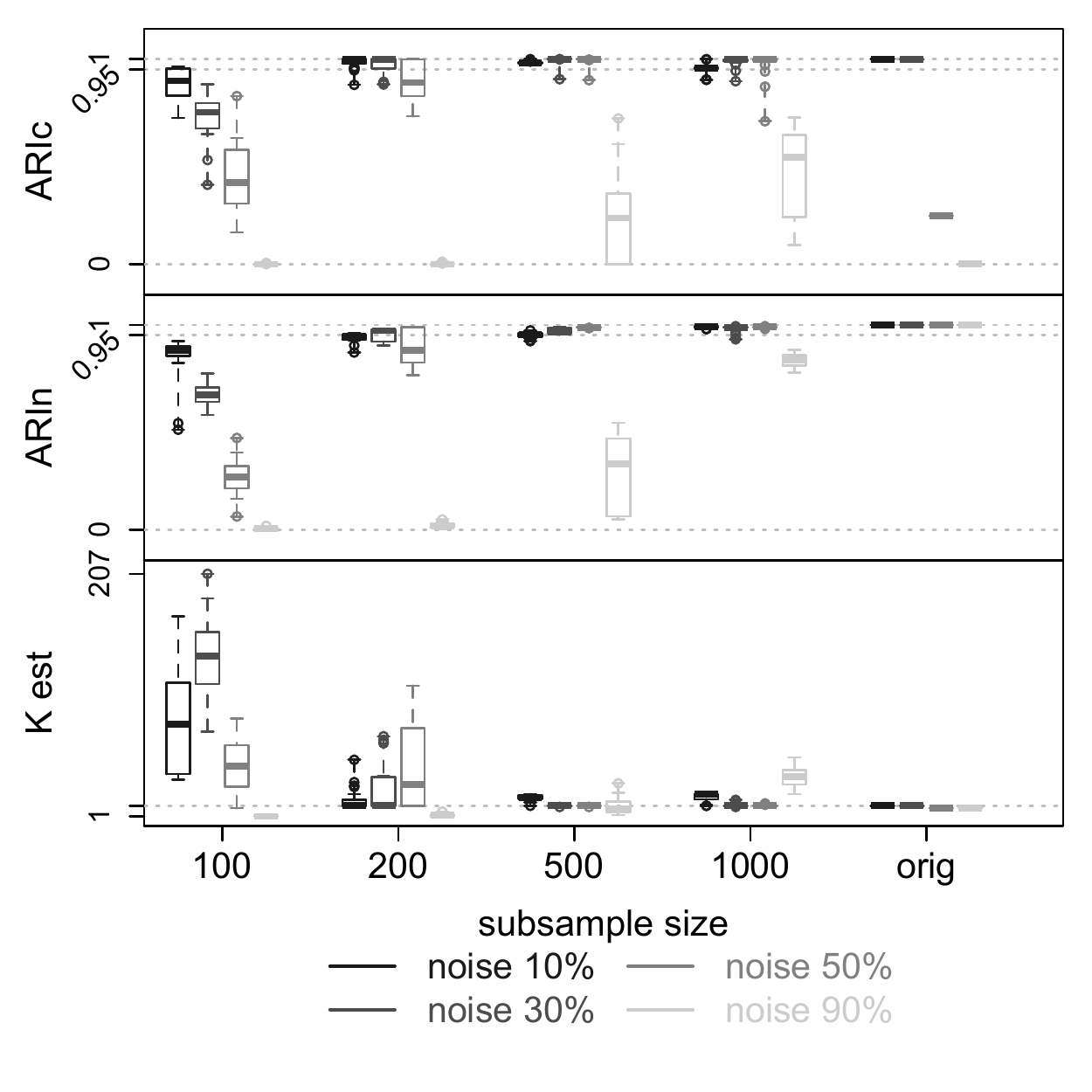}
		\caption{$n=10,000$}
		\label{ari10k_trimmed}
	\end{subfigure} %
	\begin{subfigure}[b]{0.45\textwidth}
		\centering
		\includegraphics[width=\textwidth]{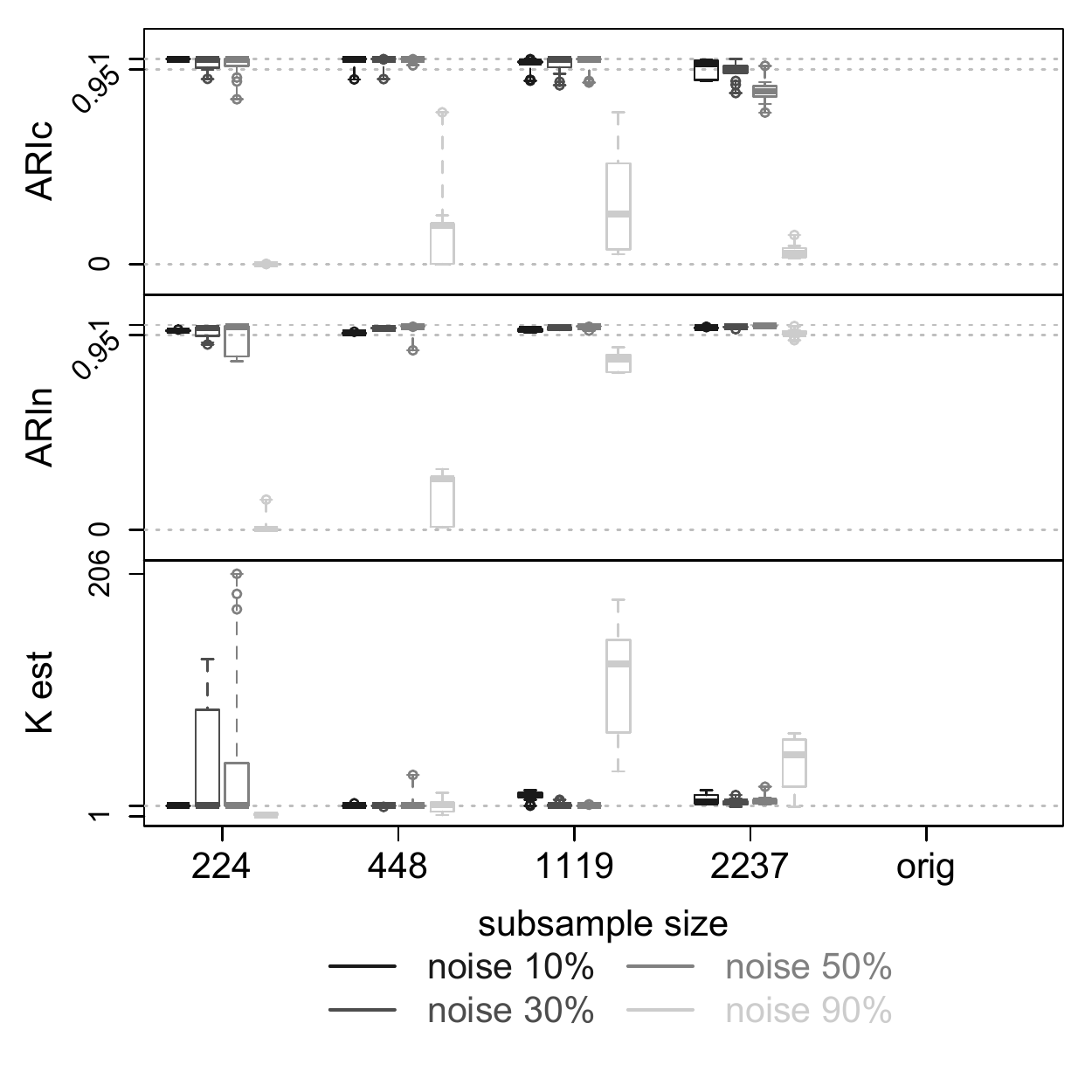}
		\caption{$n=50,000$}
		\label{ari50k_trimmed}
	\end{subfigure} %
	\caption{ARI scores and the number of estimated clusters for 10\% mean and variance trimming.}
	\label{ari_trimmed}
\end{figure*}

Overall, in these examples ISSPC shows very strong results when the noise proportion is no more than $50\%$, especially considering the small sizes of the subsamples. These results, even though somewhat inferior as expected, are quite competitive compared to the original SPC, SWIFT and even mclust. In fact, the results for smaller subsamples, $\nu=\sqrt{n}$ and $2\sqrt{n}$, have high and stable ARI scores across all the scenarios. There are several outliers in the ARI scores, indicating that the quality of the clustering result may depend on a particular random subsample, especially when there is a larger amount of noise. $\text{ARI}_c$ scores for larger subsamples, $\nu=5 \sqrt{n}$ and $10\sqrt{n}$, become more volatile, and the clustering results vary more in their quality and are generally somewhat inferior in $\text{ARI}_c$, compared to those with the smaller subsamples.
There are at least two reasons for this observation. First, the SPC solution for a larger subsample tends to have a higher variance. Second,
the use of larger subsamples increases the risk of creating, in later iterations of ISSPC, relatively large clusters that consist mostly or exclusively of noise but are accepted by the cluster selection procedure due to their substantial sizes. Thus, a smaller subsample may be beneficial not only for computational savings but also for the quality of the clustering result. Furthermore, it is seen that ISSPC does not show consistently good performance with the highest proportion of noise tested, $90\%$. However, for a suitable subsample size somewhere between 600 and 1,600, it can give a very satisfactory result, close to that of mclust, for such amount of noise, as seen from the ARI scores and the estimated 
number of clusters in Figure~\ref{ari} (the lightest gray color).
With a relatively small subsample size, ISSPC was able to provide a fairly accurate estimated number of clusters for all the  scenarios in Figure~\ref{ari}, with the exception of the cases of the highest proportion of noise, which were discussed above. 
However, ISSPC tends to overestimate the number of clusters when the subsample size is larger as a result of splitting
clusters. SWIFT has a similar tendency to split the clusters in these simulated scenarios, and the splitting is more severe, resulting in 
a very high number of estimated clusters, which is most probably due to the fact that SWIFT is generally geared towards isolating very small clusters 
in big datasets. Even with a smaller initial estimate of the number of clusters $K_0$, SWIFT still tended to assign the data points into a relatively
large number of clusters, usually greater than 20. The SWIFT algorithm also assigned more noise into the true clusters, and the need to identify the 
``background" clusters added another challenge to the accuracy comparison.

In summary, we see competitive performance of ISSPC with relatively small subsample sizes, 
compared to the results of SPC, mclust and SWIFT, as long as the noise proportion in the data is not too high ($\leq 50\%$). 
On the other hand, for such small subsample sizes, ISSPC is up to two orders of magnitude faster as shown
in the previous subsection.

\subsection{Trimming}

\begin{figure*}
	\centering
	\begin{subfigure}[b]{0.45\textwidth}
		\centering
		\includegraphics[width=\textwidth]{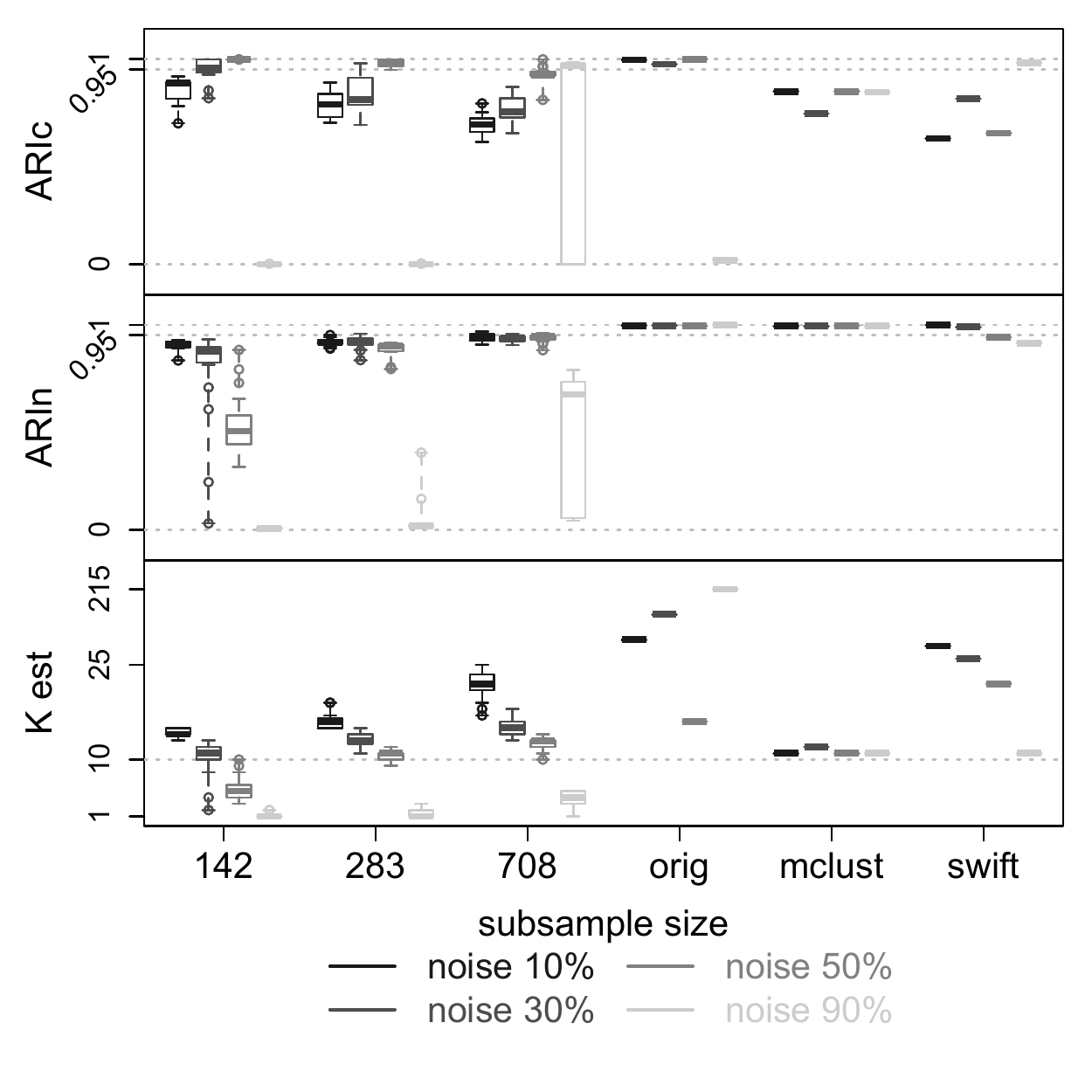}
		\caption{$n=20,000$}
		\label{ari20k_non}
	\end{subfigure} %
	\begin{subfigure}[b]{0.45\textwidth}
		\centering
		\includegraphics[width=\textwidth]{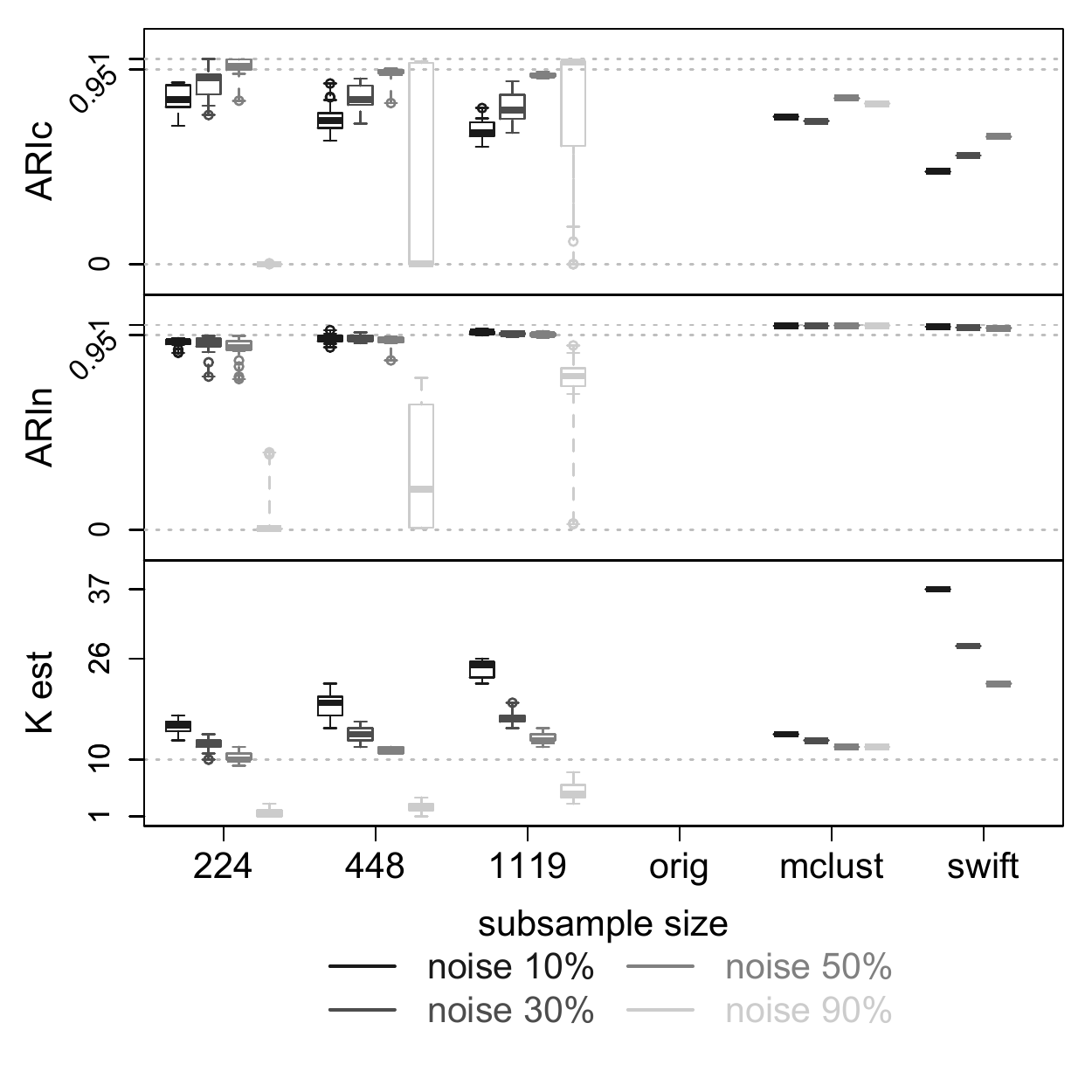}
		\caption{$n=50,000$}
		\label{ari20k_non}
	\end{subfigure} %

	\caption{ARI scores and the number of estimated clusters for non-spherical simulated clusters.
	The subsample sizes are $\sqrt{n}$, $2\sqrt{n}$, and $5\sqrt{n}$.}
	\label{ari_non}
\end{figure*}

As mentioned in Section~\ref{sec:tuning}, trimmed estimates of cluster means $\mu_k$'s and variances $\Sigma_k$'s can be useful if the clusters significantly deviate from the spherical shape, are not well separated, or generally if data are complex and high-dimensional. These situations are certainly typical for real data. Our simulated data have a simplified structure; however, to understand its potential impact, we have applied 10\% trimming to the simulated data with $n=10,000$ and $n=50,000$. 

Figure~\ref{ari_trimmed} shows boxplots of ARI scores in the top and middle panels and the estimated number of clusters in the bottom panel for both sizes. When trimming is applied to the simulated data, the clustering results are inferior to the results in Figure~\ref{ari} for smaller subsamples. It is mainly due to the fact that with trimming the clusters tend to be split, especially for clusters of a smaller size. Specifically, trimming tends to remove valid clustered data points for the calculation of the centers and variances and decreases the likelihood ratios in the sequential assignment, which results in some clustered data points being assigned to the background model. These data then create their own clusters in the subsequent clustering and sequential assignment steps. This problem is particularly severe for smaller subsample sizes,
in which cases the clusters generated by the clustering steps are necessarily small.
As the subsample size increases or the size of the clusters becomes larger, as in Figure~\ref{ari50k_trimmed}, the results improve considerably. It should also be mentioned that trimming in fact improves the results for larger subsample sizes, $\nu > 1000$, compared to those without trimming in Figures~\ref{ari10k} and \ref{ari50k}. Generally, greater amounts of trimming will produce a larger number of smaller clusters. While unfavorable for some of the simulated data scenarios, trimming can be beneficial for obtaining compact, small clusters in large datasets, which we will show in Section~\ref{sec:genedata}.

\subsection{Non-spherical clusters}
\label{sec:nonspherical}

Finally, we demonstrate the performance of ISSPC on data with non-spherical clusters. For this purpose we generated $K=10$  clusters with $p=20$ correlated dimensions for $n=20,000$ and $n=50,000$. Four out of 10 clusters were generated with 0.5 correlation, and the rest of the clusters had a correlation of 0.3. The four clusters with the highest correlation were also the largest clusters with sizes $\approx (0.3n, 0.2n, 0.1n, 0.1n)$, and the remaining six clusters were smaller and of size $\approx 0.05n$. The noisy data points were added in the same way as in the spherical data scenarios and in the same proportions. As previously, we ran ISSPC 20 times independently on each non-spherical scenario and report the average ARI scores and their distributions in boxplots in Figure~\ref{ari_non}. We again include the comparison with the original SPC algorithm, mclust and SWIFT.  

Overall, ISSPC gives a satisfactory result with non-spherical clusters, except for the cases with 90\% of noise, which still prove to be challenging. As can be expected, the non-spherical clusters tend to be split into smaller clusters. $\text{ARI}_c$ and the estimated number of clusters in Figure~\ref{ari_non} indicate that the splitting creates several very small clusters; however, the majority of the ``true" clusters are preserved correctly, albeit with a very small amount of misclassified noise. It can also be seen from the figure that mclust's and SWIFT's results are quite comparable to those of ISSPC. 
Particularly, the $\text{ARI}_c$ scores of ISSPC and mclust are close for the datasets with less than 90\% of noise, while SWIFT has somewhat lower $\text{ARI}_c$ scores. Mclust recognizes noisy data points with slightly higher accuracy, indicated by slightly better $\text{ARI}_n$ scores, mainly due to the $K$th nearest neighbor cleaning method, but the largest clusters are also split as seen from $\text{ARI}_c$ scores and the estimated number of clusters. SWIFT shows a similar tendency to produce a large number of split clusters, but it separates the remaining noise into ``background" clusters well, even for high percentages of noise. Lastly, SWIFT failed to generate any result
for the case of $n=50,000$ with 90\% of noise. 

We believe that ISSPC is very useful for data with non-spherical clusters or data that violate Gaussian assumptions as it can provide a competitive result with significant computational savings, and it is able to deliver this result for large datasets that otherwise would not be tractable for regular full data approaches such as mclust. It is feasible to modify the SPC and ISSPC algorithms in order to directly handle dependent data with a correlated structure within a cluster. Such modifications may be implemented by introducing a covariance matrix in the $\ell_2$ loss of \eqref{eq:kloss}. While it would be easy to incorporate a diagonal covariance matrix, the modifications to the algorithm would be extensive if such a covariance matrix has nonzero off-diagonal elements. A full covariance matrix would also pose further computational challenges. Even though we believe our current method might still be practical for dependent data as also shown in Section~\ref{sec:genedata}, such generalizations would be an interesting and valuable future research topic.

\section{Gene expression data}
\label{sec:genedata}

We applied ISSPC to two gene expression datasets. We first analyzed a mouse embryonic stem (ES) cell dataset \cite{zhou2007} (Zhou dataset), which was analyzed previously in \cite{marchetti2014} with the original SPC algorithm. The full dataset consists of about 45,000 genes across 16 experimental conditions; however, we used only a subset of this dataset, obtained in the same way as in \cite{marchetti2014}. The subset consists of $n=5,765$ genes in $K=2$ large distinct clusters ($1,325$ genes in the Oct4$+$ and $1,440$ in the Oct4$-$ clusters) with $3,000$ randomly selected genes whose profiles were perturbed such that these genes can be considered noise. This abridged version of the Zhou dataset is studied here as it combines the complexity of real data with a simplified noise structure. The distinct grouping patterns and the availability of information about the involved clustered genes make the clustering result easier to evaluate. The second dataset \cite{ivanova2006} (Ivanova dataset) consists of 45,264 gene expression profiles generated under different treatments in mouse ES cells across 70 experimental conditions. While gauging the behavior of ISSPC on clean, noncomplex data with the first dataset, we chose the second dataset for testing its ability to handle a full, noisy gene expression dataset. 

\subsection{Results for the Zhou dataset}
\label{sec:zhou_data}

\begin{figure*}
	\centering
	\begin{subfigure}[b]{0.25\textwidth}
		\centering
		\includegraphics[width=\textwidth]{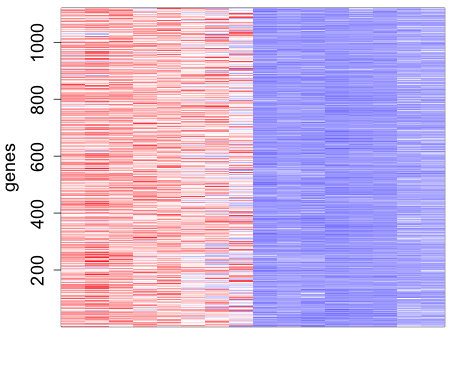}
		\caption{$1,120$ genes}
		\label{zhou_data1}
	\end{subfigure} %
	\begin{subfigure}[b]{0.25\textwidth}
		\centering
		\includegraphics[width=\textwidth]{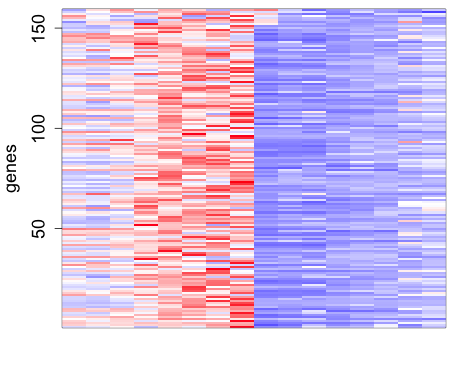}
		\caption{$159$ genes}
		\label{zhou_data2}
	\end{subfigure} \\%
	\vspace{15pt}
	\begin{subfigure}[b]{0.25\textwidth}
		\centering
		\includegraphics[width=\textwidth]{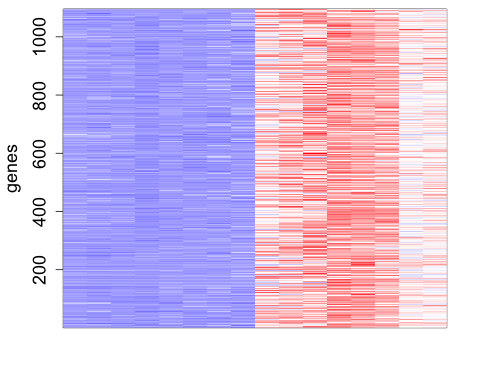}
		\caption{$1,095$ genes}
		\label{zhou_data3}
	\end{subfigure} %
	\begin{subfigure}[b]{0.25\textwidth}
		\centering
		\includegraphics[width=\textwidth]{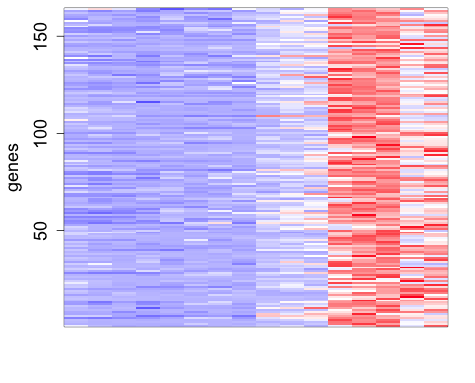}
		\caption{$164$ genes}
		\label{zhou_data4}
	\end{subfigure} %
		\begin{subfigure}[b]{0.25\textwidth}
		\centering
		\includegraphics[width=\textwidth]{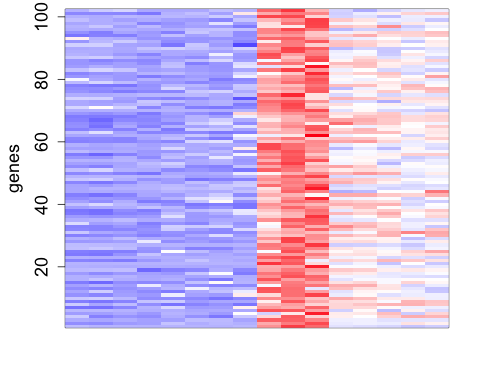}
		\caption{$102$ genes}
		\label{zhou_data5}
	\end{subfigure} %
	\caption{Five clusters obtained from the Zhou dataset. The caption of each plot indicates the size of the cluster. The experimental conditions
	1-16 are located along the $x$-axis. Blue color indicates low expression and red indicates high expression.}
	\label{zhou_data}
\end{figure*}

The Zhou dataset was run with $\eta=5$ and the subsample size $\nu = 5\sqrt{n} = 380$ as the full data size was relatively small. If ISSPC is applied to the Zhou data without the trimming as per Section~\ref{sec:tuning}, then only the two largest Oct4$+$ and Oct4$-$ clusters are found with just 4 random genes incorrectly included in these clusters and 174 Oct4 genes erroneously excluded as random. We then applied 10\% trimming and obtained five clusters with distinct grouping patterns as shown in Figure~\ref{zhou_data} in under a minute of run time. The two largest clusters in Figure~\ref{zhou_data1} and Figure~\ref{zhou_data3} recover the large Oct4$+$ and Oct4$-$ groups, respectively. There were 10 misclassified random genes and 138 misclassified Oct4 genes in these two clusters. 
The other three clusters (Figures~\ref{zhou_data2},~\ref{zhou_data4}, and~\ref{zhou_data5}) 
are relatively small and are seen to
have some subtle differences in expression patterns from the two big clusters. 
To check whether these clusters are functionally distinct from the Oct4+ and Oct$-$ groups, we performed Gene Ontology (GO) term enrichment analysis and compared the three small clusters to the full set of Oct4$+$ or Oct4$-$ genes. We found that the small cluster in Figure~\ref{zhou_data4} had several significant terms with an FDR of less than 10\% (Table~\ref{table:GO_clust4}), confirming that genes in this cluster are indeed involved
in distinct biological processes.

\begin{table*}
\begin{footnotesize}
\begin{center}
\caption{Top GO terms enriched in the small cluster shown in Figure~\ref{zhou_data4}}
\begin{tabular*}{1\textwidth}{@{\extracolsep{\fill}}m{6.5cm}m{1cm}m{1cm}m{1.5cm}m{1cm}}
\hline
  GO term & \% in cluster & \% in full set & P-value & FDR(\%)  \\ 
  \hline \\ [-0.6em]
 olfactory bulb development & 3.7 & 0.5 & $1.2 \times 10^{-5}$ & 0.0  \\ 
 olfactory lobe development & 3.7 & 0.5 & $1.2 \times 10^{-5}$ & 0.0 \\
negative regulation of Wnt signaling pathway & 6.2 & 1.5 & $4.7 \times 10^{-5}$ & 2.7 \\
embryonic organ morphogenesis & 8.7 & 3.2 & $2.7 \times 10^{-4}$ & 7.0 \\
central nervous system development & 13.0 & 6.2 & $4.6 \times 10^{-4}$ & 7.2 \\
 regulation of Wnt signaling pathway & 7.5 & 2.7 & $6.9 \times 10^{-4}$ & 10.0 \\ [0.2em]
    \hline
\end{tabular*}
\label{table:GO_clust4}
\end{center}
\end{footnotesize}
\end{table*}

Although the two clusters in Figures~\ref{zhou_data2} and~\ref{zhou_data5} did not have any GO terms with FDR $< 10\%$, 
it can be clearly seen from the figures that they both have distinct expression patterns from the other clusters.
The cluster in Figure~\ref{zhou_data2} has much lower expression levels in the first three conditions and somewhat higher levels in conditions 6-8 compared to the large Oct4$+$ cluster in Figure~\ref{zhou_data1}. The cluster in Figure~\ref{zhou_data5} has a particularly high expression pattern across conditions 9-11, compared to all the other clusters. Detecting such heterogeneity may be useful for novel findings from a big dataset.

\subsection{Results for the Ivanova dataset}
\label{sec:ivanova_data}

\begin{figure*}
	\centering
	\begin{subfigure}[b]{0.22\textwidth}
		\centering
		\includegraphics[width=\textwidth]{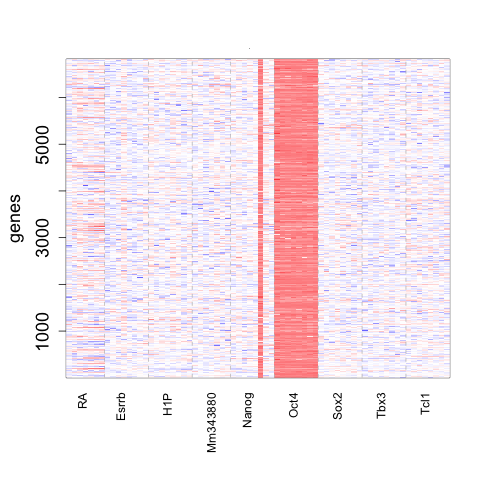}
		\vspace{-20pt}
		\caption{6,823 genes}
		\vspace{-5pt}
		\label{iva_data1}
	\end{subfigure} %
		\begin{subfigure}[b]{0.22\textwidth}
		\centering
		\includegraphics[width=\textwidth]{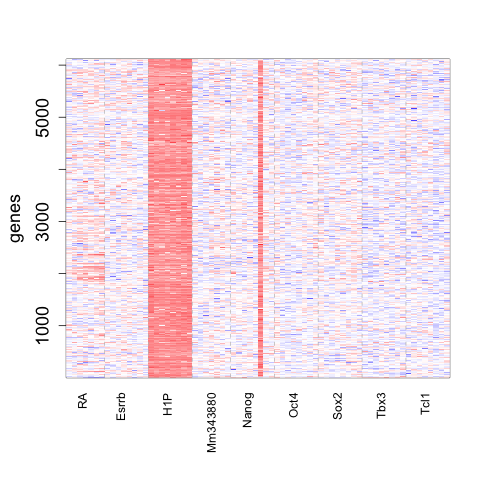}
		\vspace{-20pt}
		\caption{6,116 genes}
		\vspace{-5pt}
		\label{iva_data2}
	\end{subfigure} \\%
	\begin{subfigure}[b]{0.22\textwidth}
		\centering
		\includegraphics[width=\textwidth]{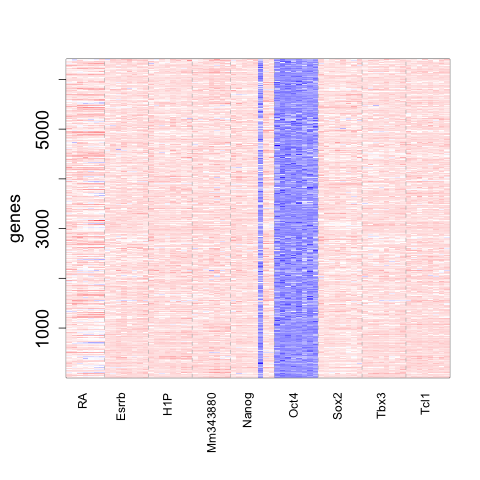}
		\vspace{-20pt}
		\caption{6,409 genes}
		\vspace{-5pt}
		\label{iva_data3}
	\end{subfigure} %
		\begin{subfigure}[b]{0.22\textwidth}
		\centering
		\includegraphics[width=\textwidth]{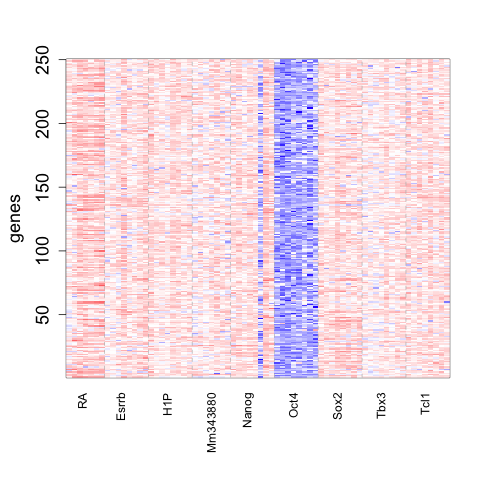}
		\vspace{-20pt}
		\caption{250 genes}
		\vspace{-5pt}
		\label{iva_data4}
	\end{subfigure} %
	\begin{subfigure}[b]{0.22\textwidth}
		\centering
		\includegraphics[width=\textwidth]{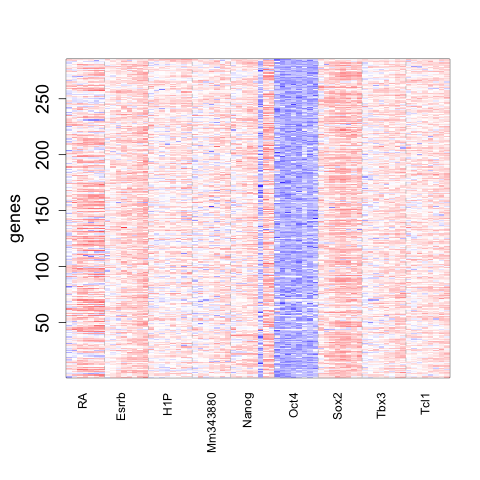}
		\vspace{-20pt}
		\caption{285 genes}
		\vspace{-5pt}
		\label{iva_data5}
	\end{subfigure}\\ %
		\begin{subfigure}[b]{0.22\textwidth}
		\centering
		\includegraphics[width=\textwidth]{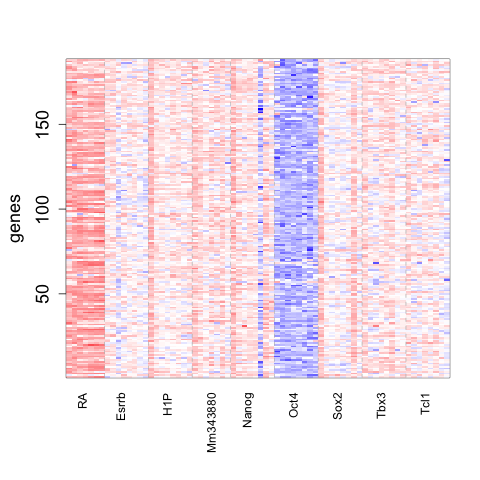}
		\vspace{-20pt}
		\caption{188 genes}
		\vspace{-5pt}
		\label{iva_data6}
	\end{subfigure} %
	\begin{subfigure}[b]{0.22\textwidth}
		\centering
		\includegraphics[width=\textwidth]{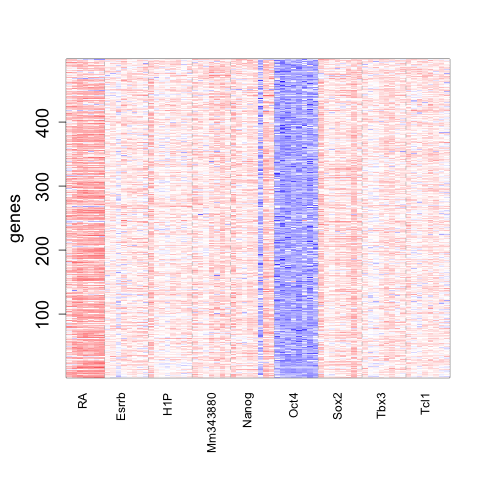}
		\vspace{-20pt}
		\caption{498 genes}
		\vspace{-5pt}
		\label{iva_data7}
	\end{subfigure} %
	\begin{subfigure}[b]{0.22\textwidth}
		\centering
		\includegraphics[width=\textwidth]{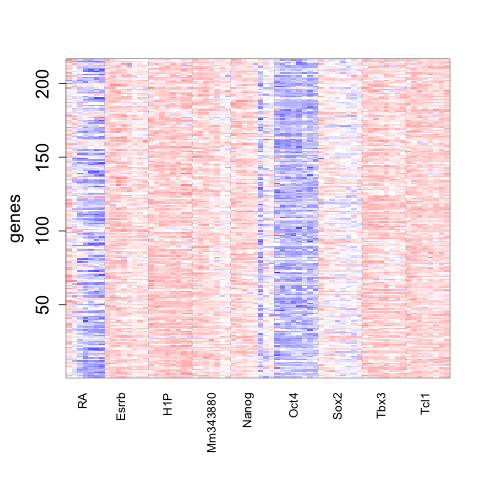}
		\vspace{-20pt}
		\caption{216 genes}
		\vspace{-5pt}
		\label{iva_data8}
	\end{subfigure}\\ %
	\begin{subfigure}[b]{0.22\textwidth}
		\centering
		\includegraphics[width=\textwidth]{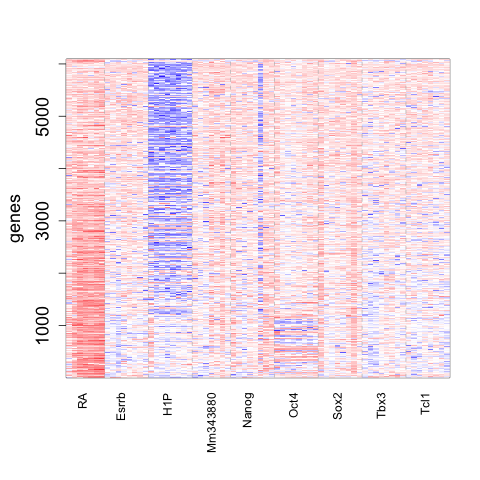}
		\vspace{-20pt}
		\caption{6,093 genes}
		\vspace{-5pt}
		\label{iva_data9}
	\end{subfigure} %
	\begin{subfigure}[b]{0.22\textwidth}
		\centering
		\includegraphics[width=\textwidth]{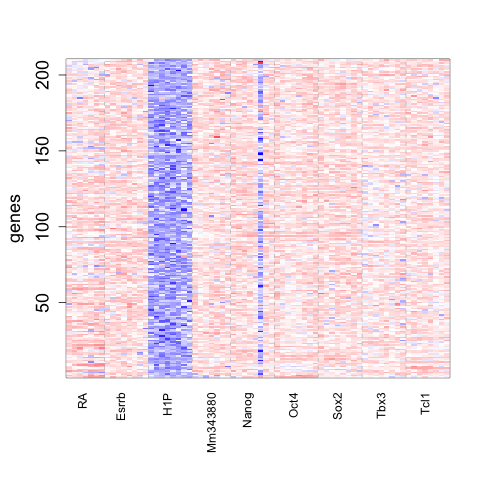}
		\vspace{-20pt}
		\caption{210 genes}
		\vspace{-5pt}
		\label{iva_data10}
	\end{subfigure} %
	\begin{subfigure}[b]{0.22\textwidth}
		\centering
		\includegraphics[width=\textwidth]{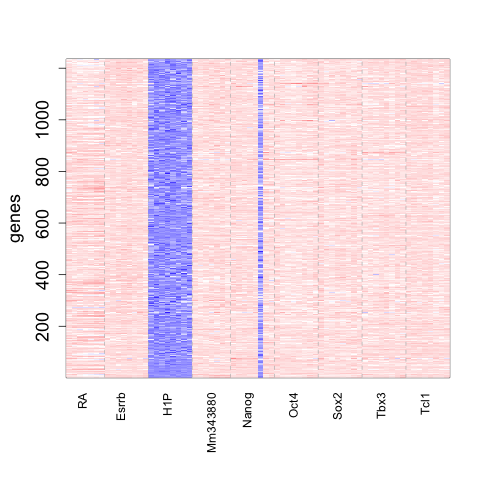}
		\vspace{-20pt}
		\caption{1,235 genes}
		\vspace{-5pt}
		\label{iva_data11}
	\end{subfigure}\\ %
	\begin{subfigure}[b]{0.22\textwidth}
		\centering
		\includegraphics[width=\textwidth]{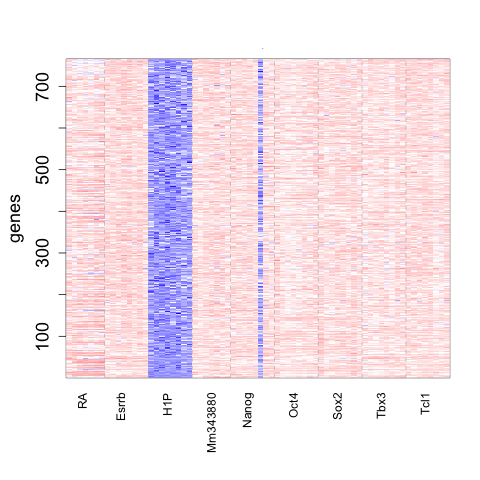}
		\vspace{-20pt}
		\caption{765 genes}
		\vspace{-5pt}
		\label{iva_data12}
	\end{subfigure} %
	\begin{subfigure}[b]{0.22\textwidth}
		\centering
		\includegraphics[width=\textwidth]{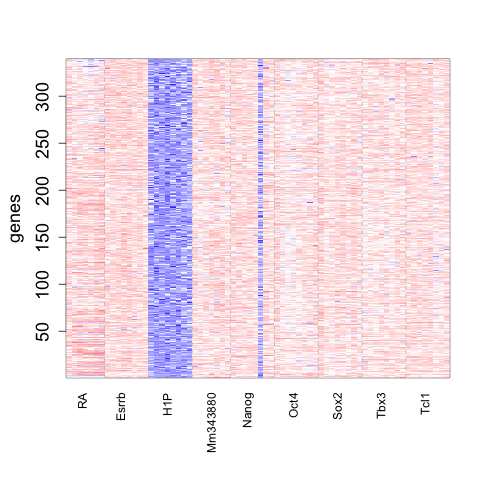}
		\vspace{-20pt}
		\caption{339 genes}
		\vspace{-5pt}
		\label{iva_data13}
	\end{subfigure} %
	\caption{Clusters obtained from the Ivanova dataset. From left to right the experimental conditions are labeled as RA, Esrrb, 
	control condition (H1P), Mm343880, Nanog, Oct4, Sox2, Tbx3, Tcl1.}
	\label{iva_data}
\end{figure*}

The gene expression data in \cite{ivanova2006} were generated under retinoid acid (RA) induction, a control condition, and the knockdown experiments of seven transcription factors (Oct4, Nanog, Sox2, Esrrb, Tbx3, Tcl1 and Mm343880) over approximately eight days to explore the mechanisms of self-renewal and differentiation of mouse ES cells. The study in \cite{ivanova2006} clustered 3,109 genes, and another study in \cite{mason2009} clustered about 17,000 genes from this dataset. We attempt to cluster the full dataset of 45,264 genes without any filtering to identify more potential patterns. The ISSPC algorithm was applied with $\eta=15$, 20\% trimming, and subsample size $\nu = 2 \sqrt{n} = 425$. We obtained 13 clusters of various sizes, depicted in Figure~\ref{iva_data}, and identified a total of about 15,000 genes as noise. Hereafter, we characterize a cluster from this dataset as condition-high or condition-low, for example, if a cluster is described as Oct4-high, then it means that genes in this cluster have high expression in the Oct4 knockdown condition. Overall, the obtained clusters display four main distinct expression patterns: high expression in Oct4 knockdown in Figure~\ref{iva_data1}, high expression in the control condition (H1P) in Figure~\ref{iva_data2}, low expression in Oct4 knockdown in Figures~\ref{iva_data3}-\ref{iva_data8}, and low expression in H1P in Figures~\ref{iva_data9}-\ref{iva_data13}. While the first two patterns are represented by two large clusters of sizes approximately 6,800 and 6,100 genes, respectively, each of the latter two is separated into one big cluster of size $>6,000$ and a few smaller groupings. We again turn to GO term enrichment analysis to determine whether the clusters, particularly the small clusters with low expression in Oct4 knockdown, contain any biologically relevant genes.   
\begin{table*}
\begin{footnotesize}
\begin{center}
\caption{Top unique GO terms enriched in the two clusters from Figure~\ref{iva_data1} and Figure~\ref{iva_data3}}
\begin{tabular*}{1\textwidth}{@{\extracolsep{\fill}}m{0.4cm}m{6cm}m{1cm}m{1cm}m{1.5cm}m{1cm}}
\hline
 cluster & GO term & \% in cluster & \% in full set & P-value & FDR(\%)  \\ 
  \hline \\ [-0.6em]
 \ref{iva_data1} & regulation of localization & 13.2 & 8.3 &  9.2e-47 & 0.0 \\ 
  & regulation of transport & 10.1 & 6.1 &  7.2e-41 & 0.0 \\ 
 & cell surface receptor signaling pathway & 15.3 & 10.4 &  2.7e-39 & 0.0 \\ 
  \\
& receptor binding & 9.5 & 5.7 &  1.3e-38 & 0.0 \\ 
 & receptor activity & 7.4 & 4.6 &  2.2e-27 & 1.6 \\ 
 & signal transducer activity & 7.8 & 4.9 &  1.1e-26 & 1.3 \\  [0.2em]
    \hline \\ [-0.6em]
\ref{iva_data3} &  cellular response to DNA damage stimulus & 5.1 & 2.7 &  2.9e-28 & 0.1 \\ 
 & cellular macromolecular complex assembly & 4.6 & 2.5 &  1.0e-23 & 0.3 \\ 
 & mitochondrion organization & 2.8 & 1.3 &  1.5e-22 & 0.4 \\ 
  \\
&  structural constituent of ribosome & 2.1 & 0.5 &  9.1e-48 & 0.0 \\ 
& hydrolase activity & 14.7 & 10.3 &  2.3e-29 & 0.1 \\ 
& hydrolase activity, acting on acid anhydrides & 5.6 & 3.1 &  8.1e-27 & 0.4 \\  [0.2em]
   \hline
\end{tabular*}
\label{table:GO_bigoct4}
\end{center}
\end{footnotesize}
\end{table*}

GO term enrichment analysis indicates that all of the discovered clusters are in fact biologically relevant, containing genes enriched in both function and process categories. All of the clusters, however, shared many broad terms such as ``binding" or ``cellular process" with a large number of genes involved in these high level functions and processes, and thus, we report only top unique GO terms in each cluster. The large H1P-low cluster in Figure~\ref{iva_data9} as well as any of the smaller H1P-low clusters in Figures~\ref{iva_data10}-\ref{iva_data13} did not have any unique GO terms associated with them. The H1P-high cluster in Figure~\ref{iva_data2} contained approximately one third of genes in the ``unannotated" process and function categories with $P=4.8 \times 10^{-24}$ and FDR $\approx 6\%$, which means that the roles of these genes have not yet been annotated and/or not yet determined. The six significant unique GO terms for the Oct4-high and -low clusters are presented in Table~\ref{table:GO_bigoct4}.

All of the small Oct-low clusters (Figures~\ref{iva_data4}-\ref{iva_data8}) have markedly evident expression differences from the big Oct4-low cluster of Figure~\ref{iva_data3}. Four out of the five small Oct4-low clusters were enriched for multiple  unique GO terms with FDR $< 5\%$, mostly in the process ontology, confirming that these small clusters are biologically relevant and might have some distinct biological roles. Some examples of the significant unique GO terms of these clusters included reproductive system development, heart development, and cell cycle phase transition. A table of the associated unique significant GO terms is provided in the supplemental materials. Among these small Oct4-low clusters the one in Figure~\ref{iva_data5} is particularly interesting. In addition to low expression when Oct4 is knocked down, this cluster
shows a high expression pattern in Sox2 knockdown, using the expression level in the control condition (H1P) as a reference. Thus, the two transcription factors, Oct4 and Sox2, regulate the genes in this cluster in an opposite way.
This finding is in sharp contrast to the established co-regulation roles between Oct4 and Sox2 in ES cells 
(\cite{zhou2007} and references therein), often regulating genes in a coordinated way by binding to adjacent DNA sites
\cite{mason2010}. This cluster was uniquely enriched for genes responsible for reproductive system development. Another class of GO terms that were significant for this cluster can be classified as relating to cardiac and heart system development, all with $P \leq 5.1 \times 10^{-4}$ and FDR $<5\%$.  

It appears that none of the H1P-low clusters, including the largest one in Figure~\ref{iva_data9}, contained any unique GO terms, and moreover, none of the small H1P-low clusters were enriched for genes with a significant GO terms when compared to the big H1P-low cluster.  
It is possible that these clusters are a result of splitting as experienced with the non-spherical simulated data. Gene expression data usually contain clusters with a correlated structure, and splitting might be compounded due to a high degree of trimming for this particular cluster. As a result, splitting of some big clusters might be the price of discovering tight clusters in such a dataset. Fortunately, split clusters can be easily detected and merged, especially given a reasonably small number of them. A possible remedy is to decrease the amount of trimming. For example, if 10\% trimming is applied to the clusters before the sequential assignment step, then 8 clusters will be obtained with the four largest clusters preserved. The other  small clusters include one with the same pattern as in Figure~\ref{iva_data8}, one similar to that in Figure~\ref{iva_data7}, and two small H1P-low clusters instead of four. The smaller amount of trimming thus created larger clusters with less splitting, but some special patterns were not extracted from the big Oct4-low cluster.                    

The results on the Ivanova dataset have shown that ISSPC is able to obtain clusters that have meaningful biological functions. Moreover, some of the obtained clusters were very small, of sizes between 200 and 500. These small clusters were extracted from a dataset of over 45,000 genes along with clusters of size approximately 6,000, demonstrating the ability of ISSPC to find tight clusters with very small subsamples. These results were achieved in about 30 minutes of computational time and would not be computationally feasible with the original SPC algorithm. The computational savings of the iterative subsampling approach make it feasible to explore this dataset even further by varying the degree of trimming or the dimension number cutoff $\eta$ and to discover more interesting patterns.

\section{Discussion}

We believe that ISSPC is a very promising approach to clustering as it allows us to perform computation with a relatively low trade off between speed and accuracy. It can successfully cluster large datasets in a reasonable amount of time. Such large datasets would otherwise require either filtering or some other preprocessing, which could potentially remove some valuable data points. Gene expression datasets, for instance, are typically filtered based on coefficient of variation, and the genes that do not make a certain coefficient of variation cutoff are removed. Thus, some genes that do in fact carry patterns in their expression profiles and play a role in certain biological functions can be lost and could confound the analysis. We have shown that ISSPC is able to produce adequate results with real data. Moreover, it generates a solution without the prior knowledge of the number of clusters and is able to effectively separate the noise.           

The ISSPC algorithm requires some fine tuning, which involves the dimension number cutoff $\eta$, the trimming percentage, and the subsample size $\nu$. The solution can be sensitive to each of these tuning parameters. The computational savings, however, make it possible to rerun a dataset with various values of these parameters. Additionally, 
in light of the problem of split clusters, especially with trimming, it is possible to address and introduce a merging step after the recursions are terminated. Such a merging step would make it easier for the user to interpret the clustering result and can potentially help with separating overlapping versus truly homogeneous clusters. 

Iterative subsampling with sequential assignment can be applied generally to accelerate any advanced clustering method that has inferior time complexity, such as mclust, PWK-means \cite{tseng2007}, convex clustering \cite{chi2013}, or PRclust \cite{pan2013}. However, the clustering mechanism at the center of the iterative subsampling approach would need to be able to estimate the number of clusters or provide a solution path as well as to be able to isolate noisy data points. The simplicity of implementation and easy intuition alongside the orders of magnitude of computational savings could present further appeal of this methodology for big data.

\appendix
\section*{Appendix}

Let $h(k) = \left( \begin{array}{c} k \\ 2 \end{array} \right)$. Given a contingency table $(n_{ij} )_{I \times J}$ with entries $n_{ij}$, row sums $n_{i \bullet} = \sum_{j} n_{ij}$, column sums $n_{\bullet j } = \sum_{i} n_{ij}$, and a total sum of entries $n = \sum_{i,j} n_{ij}$, let $h_1=\sum_{i} h( n_{i \bullet})$ and $h_2=\sum_{j} h(n_{\bullet j})$. Then the ARI is defined by 
\begin{align}
\label{eq:ari}
\text{ARI} = \frac{\sum_{i,j} h \left( n_{ij} \right) -  h_1 h_2   /   h (n)} {\frac{1}{2}(h_1+h_2)-  h_1 h_2 / h(n)}.
\end{align}

Suppose the true partition and an estimated one is 
\begin{equation*}
C = \left\{ C_1, \ldots, C_R, C_{R+1}\right\} \text{ and } \hat{C} = \{ \hat{C}_1, \ldots, \hat{C}_K, \hat{C}_{K+1} \}, 
\end{equation*}
respectively, where $C_r$ and $\hat{C}_k$ contain the indices of the data points assigned to true and estimated clusters for $r=1, \ldots, R$ and $k=1, \ldots, K$, respectively, and $C_{R+1}$ and $\hat{C}_{K+1}$ are the indices assigned to noise. Denote the respective cluster labels as $v = \left\{ v_1, \ldots, v_R, v_{R+1} \right\}$ and $\hat{u} = \left\{ \hat{u}_1, \ldots, \hat{u}_K, \hat{u}_{K+1} \right\}$, where $v_{R+1}$ and $\hat{u}_{K+1}$ indicate the labels for the true and estimated noise, respectively. Table~\ref{table:aricounts} shows the full contingency table of the counts $n_{kr} = |\hat{C}_k \cap C_r |$. Our ARI scores are calculated based on parts of the counts in this table. 

\begin{table}
\begin{center}
\caption{Contingency table and notation for the calculation of $(\text{ARI}_c,\text{ARI}_n)$}
\begin{tabular*}{0.49\textwidth}{@{\extracolsep{\fill}}c|cccc|c}
Cluster & $v_1$ & \ldots & $v_R$ & $v_{R+1}$ & Sum \\
\hline
$\hat{u}_1$ & $n_{11}$  & \ldots & $n_{1R}$ &  $n_{1(R+1)}$ & $n_{1\bullet}$ \\
\vdots & \vdots  & \vdots & \vdots & \vdots  & \vdots \\
$\hat{u}_K$ & $n_{K1}$ & \ldots & $n_{KR}$ & $n_{K(R+1)}$ &  $n_{K\bullet}$ \\
$\hat{u}_{K+1}$ & $n_{(K+1)1}$ & \ldots & $n_{(K+1)R}$ & $n_{(K+1)(R+1)}$ & $n_{(K+1)\bullet}$ \\
\hline
Sum & $n_{\bullet1}$ & \ldots & $n_{\bullet R}$ & $n_{\bullet (R+1)}$ &  $n$ \\
\end{tabular*}
\label{table:aricounts}
\end{center}
\end{table}

$\text{ARI}_c$ accounts for data points that are identified as belonging to estimated clusters $\hat{C}_k$, $k=1, \ldots, K$, and is calculated as in \eqref{eq:ari} using only the first $K$ rows of Table~\ref{table:aricounts}, 
\begin{equation*}
\{ n_{kr}: 1 \leq k \leq K, 1 \leq r \leq R+1 \},
\end{equation*}
with the corresponding row and column sums. In effect, $\text{ARI}_c$ provides quality assessment of the identified clusters $\hat{C}_k$, i.e., misclassification of clustered data and the amount of noise in the estimated clusters.

 $\text{ARI}_n$ indicates how sensitive a method is in identifying noise and whether any clustered data point is misclassified as noise. It is based on all the data points except the noise in the estimated clusters, which is accounted for in $\text{ARI}_c$. We collapse Table~\ref{table:aricounts} to a $2 \times 2$ table with counts $n_{cc}^* = \sum_{k=1}^K \sum_{r=1}^{R} n_{kr}$, $n_{nn}^* = n_{(K+1)(R+1)}$, $n_{nc}^* = \sum_{r=1}^{R} n_{(K+1)r}$, and $n_{cn}^* = \sum_{k=1}^{K} n_{k(R+1)}$. We then set $n_{cn}^* = 0$ since we account for these data points in $\text{ARI}_c$. The total number of data points to be considered for $\text{ARI}_n$ is thus $n^* = n^*_{cc} + n^*_{nn} + n^*_{nc}$. Again, we use \eqref{eq:ari} to calculate $\text{ARI}_n$ by plugging in $n^*_{cc}$, $n^*_{nn}$, $n^*_{nc}$, and $n_{cn}^* = 0$. 
 In general, if $n^*_{nn}$ is large and $n^*_{nc}$ is small, $\text{ARI}_n$ will be close to 1.

\bibliographystyle{plainnat}
\bibliography{spcreferences}

\end{document}